\documentclass[preprint,12pt]{elsarticle}

\usepackage[margin=2.5cm]{geometry}

\usepackage{amsmath}
\usepackage{amssymb}

\journal{Journal of Computational and Applied Mathematics}

\begin{document}
\begin{frontmatter}


\title{Fast resolution of a single factor
Heath-Jarrow-Morton model with stochastic volatility}

\author{
E. Valero$^1$ ,  M. Torrealba$^2$, L. Lacasa$^{1}$ and F.
Fraysse$^1$}
\address{ $^1$Dpto. de Matem\'{a}tica Aplicada, ETSI Aeron\'{a}uticos, Universidad Polit\'{e}cnica de Madrid, Spain. \\
 $^2$Grupo BBVA, Mercados Globales y Distribuci\'on, Desarrollo de Nuevos Productos.}

\begin{abstract}
This paper considers the single factor Heath-Jarrow-Morton model
for the interest rate curve with stochastic volatility. Its
natural formulation, described in terms of stochastic differential
equations, is solved through Monte Carlo simulations, that usually
involve rather large computation time, inefficient from a
practical (financial) perspective. This model turns to be
Markovian in three dimensions and therefore it can be mapped into
a 3D partial differential equations problem. We propose an
optimized numerical method to solve the 3D PDE model in both low
computation time and reasonable accuracy, a fundamental criterion
for practical purposes. The spatial and temporal discretization
are performed using finite-difference and Crank-Nicholson schemes
respectively, and the computational efficiency is largely
increased performing a scale analysis and using Alternating
Direction Implicit schemes. Several numerical considerations such
as convergence criteria or computation time are analyzed and
discussed.
\end{abstract}


\end{frontmatter}


\section{Introduction}
\label{introduction}

In quantitative finance, the interest rate curve has been
intensely studied and modeled
 in terms of stochastic differential equations (SDE), assuming for this curve a
temporal evolution
which satisfies the non-arbitrage opportunity
in a complete and efficient market \cite{HWgeneral, rebo}. Such restriction implies that
the model at hand needs to be calibrated by the market prices of the most liquid instruments,
 bond prices being a representative example. These in turn are decomposed in more elementary units, the
so called \textit{zero coupon bonds}. These are, roughly speaking,
financial instruments that pay one unit of currency at a certain
future date (maturity). The price of these latter instruments will
characterize the average behavior of the interest rate curve
\cite{HJMseminal}. On the other hand, the fluctuations of the
interest rate curve with respect to its average, the so called
volatility, is quantified through market instruments such as
\textit{caps} and \textit{floors} \cite{general}. A cap pays the
difference between a certain rate and a certain prespecified level
(strike) if this difference is positive, while the floor pays off
the difference between the strike and the value of the rate, if
positive. The relation that links the cap or floor premium with
the volatility is the well known Black-Scholes
formula \cite{HWgeneral}.\\

Interested in the calibration of zero coupon bond prices together
with cap/floor prices, in this paper we address the
Heath-Jarrow-Morton model (HJM) which in its natural formulation
is described in \cite{HJMseminal}. The main interest of the HJM
approach is set on the fact that it provides a broad mathematical
formulation \cite{HJMgeneral}, where most of market observed
features can be taken into account. This wasn't the case of
preceding seminal models such as Black-Derman-Toy model or
Hull-White models \cite{HWgeneral}, where the model parameters
were very difficult to calibrate in terms of market observed
patterns. Furthermore, those models only incorporated one source
of stochasticity, and therefore the only possible mode in the
interest rate curve was the parallel movement. Only lognormal or
normal statistical distributions of the short rate were possible.
These distributions for the short rate arise in distributions for
different maturities of the libor rates, which are far away from
those implied in the markets. Conversely, HJM models are not
afflicted by these drawbacks. The HJM framework \cite{general} is
general, in the sense that many previous models describing the
term structure of interest rates can be understood as particular
cases of a HJM model, that in turn can incorporate as many risk
factors as needed in order to accurately describe the evolution of
the rate's curves. The formulation of the HJM can be extended to
incorporate several stochastic factors; hence, the simulated
interest rates curve movements could include deformation modes
that changes the initial slope and convexity of this curve, in
order to describe the covariance and autocorrelation structures
present in the time behaviour of the interest rate curve.
Additionally, the probability distribution function of interest
rates can be exogenously defined by means of local volatility
functions (of deterministic or stochastic nature)
in order to match option prices as quoted in the market.\\

The major drawback of HJM models is set on the continuous nature
of its state variables (the continuous time structure of forward
rates), what leads to an infinite amount of state variables
\cite{HJMgeneral}. Therefore, in general, these models are
non-Markovian in finite dimensions, and thus can only be solved by
using fitted Monte Carlo techniques which are eventually slow and
computationally delicate. This is of course problematic for
practical purposes, since financial industry requires models that
can be integrated in real time with reasonable accuracy. Quite
interestingly, HJM models can be in some particular situations
transformed into a low order Markovian system \cite{Markov1,
Markov2, Markov3, Markov4, Markov5}, and therefore can be
subsequently mapped to a partial differential system due to the
well known Feynman-Kac formula \cite{Feynman}. Numerical methods
for solving partial differential systems can consequently apply
\cite{libro general, Tavella}. Although integration of PDE is
typically faster than Monte Carlo simulations of the associated
SDE system, in practice one needs additional optimization methods
for calibration (conjugate gradient, genetic algorithms), thus
requiring a large number of evaluations of the model.
To find out fast and efficient model evaluations (numerical solvers) is therefore demanding in financial industry.\\

In this paper, we propose a full numerical methodology to optimize
such issues. Amongst the plethora of different HJM models, we
address here a \textit{single factor} HJM model with
\textit{stochastic volatility}. This choice is justified on
practical reasons (this is a realistic enough model which is
actually used in the financial industry \cite{findus1, findus2})
and can be argued as it follows: (i) \textit{single factor}:
Principal Component Analysis of the covariance matrix associated
to historic data of the interest rate curve suggest that the first
eigenvalue has a weight of $85\%$ \cite{PCA2, PCA1}. (ii)
\textit{stochastic volatility}: The preliminary HJM models used to
model the stochastic behavior of the interest rate market by a
single Wiener process that drives the forward rate processes (that
is to say, there was a single source of stochasticity in the
models). In the last years, some authors have introduced
\cite{vola}, according to empirical evidence, a new source of
stochasticity in the description of the volatility evolution,
leading to the so called stochastic volatility HJM
models \cite{stoc.vol.}.\\

Amongst all the possible stochastic volatility functions, we will
focus on those which are separable, that is, that can be
factorized in the product of a stochastic function and a time
dependent deterministic function and another deterministic
function that depends on the maturity. Again, this choice isn't
random, much on the contrary, the resulting HJM model turns to be
Markovian in only three dimensions, and can therefore be mapped
and evaluated within a PDE framework. We will present a complete
numerical integration of the model based on finite differences
schemes and Crank-Nicholson temporal integrators. In order to
drastically improve the computational efficiency of the method
(that is, reaching fast computation time while preserving good
accuracy), we employ a scale analysis for the mesh optimization
and, for the first time in financial models (as far as we are
aware), Alternating Direct Implicit (ADI) schemes, techniques
borrowed from computational fluid dynamics. The rest of the paper
goes as follows:  Section~\ref{sec:CurvaCupOnCero} focuses on the
interest rate market, defining the zero coupon bond as the trading
derivative. Then, section ~\ref{HJM} presents the single factor
HJM model with stochastic volatility. The model is Markovianised
using subsidiary state variables, as usual, and the specific
volatility function enables a 3D PDE formulation. Section
~\ref{numerical} and \ref{nummethod} describe the full numerical
method. The numerical validation is depicted in section
\ref{validation}. After some additional remarks regarding the
numerical methods (section \ref{additional}), in section
\ref{conclusion} we conclude.

\section{Interest rate market: the zero coupon bond}
\label{sec:CurvaCupOnCero}
In the interest rate market, the zero
coupon bonds are taken as the most basic market instruments in the
sense that any other financial quantity related to interest rates
can be derived from them. These assets merely pay a monetary unit
in a given future time that is called the expiration date, and the
current price as a function of the expiration date is determined
by the so called zero coupon bond curve (ZCBC). We will formally
denote this curve as it follows:
\begin{eqnarray}
p\left(t,T\right), t\in \mathbb{R}^{+}, T\geq t: T\rightarrow
p(t,\cdot),
\end{eqnarray}
where $t$ stands for the present time (valuation date), $T$ stands
for the bond expiration date and $p\left(t,T\right)$ is merely the
zero coupon bond price (ZCBP). We will also assume from now on
that ZCBC fulfills the necessary regularity conditions.
\\
By definition, ZCBC is such that:
\begin{itemize}
    \item the ZCBP that expires at the present time is 1 (trivially): $p(T,T)=1$.
    \item ZCBP $\in (0,1]$ and is a monotonically decreasing function (this property is assumed in order to avoid the existence of arbitrage).
\end{itemize}
It is worth saying that in exceptional macroeconomic situations,
it is actually possible that the latter property doesn't hold, due
to the intervention of central banks: of Japan, in order to encourage investors to buy their currency \cite{japan}.\\
Provided the preceding properties, there always exists a function
$f(t,s), s\geq t $ such that
\begin{eqnarray}
& &p(t,T) = e^{-\int^{T}_{t}f(t,s)ds}, t\in \mathbb{R}^{+}, T\geq t\nonumber\\
& &f(t,s)\geq0, \forall s\geq t
\end{eqnarray}
and consequently, $f$ fulfills:
\begin{eqnarray}
f(t,T) = - \frac{\partial \ln p(t,T)}{\partial T},\forall T\geq t.
\end{eqnarray}
>From a financial point of view, $f(t,T)$ is interpreted in terms
of the interest rate that an investor would receive if he sells in
$t$ a zero coupon bond with expiration date $T$ and buys another
with expiration date $T+dT$. This bond is the so called forward
rate of interest. In particular, the short term rate of interest
$r$ fulfills:
\begin{eqnarray}
r(t) = f(t,t) = \left. - \frac{\partial \normalfont \ln
p(t,T)}{\partial t} \right| _{T=t},
\end{eqnarray}
and is of special interest because it quotes the return of an
investment of one monetary unit in the present time, \( t\), that
is redeemed an infinitesimal time later, \(t+dt\).

\section{HJM framework}
\label{HJM}

\subsection{The model}
The single factor HJM model under hands is originally represented
by the stochastic differential equation for the short rate $r$
(details can be found in \cite{general,new})
\begin{eqnarray}
&&dr(t) =  \left(\frac{\partial f(0,t)}{\partial t} -
\kappa(r(t)-f(0,t))+y(t)\right)dt+\eta (t,r(t)) dW(t).\nonumber\\
&&dy(t) = [\eta (t,x(t))^{2} -2\kappa y(t)]dt,
\end{eqnarray}
where $y(t)$ is a subsidiary state variable with no financial
meaning, employed to Markovianise the model, $\kappa$ is a
positive constant and $dW\left(t\right)$ is a Wiener process. The
volatility function $\eta(t,r(t))$ is defined through
\begin{eqnarray}
& &\eta(t,r(t)) = \sqrt{v(t)} \lambda (t) r(t)^{\gamma (t)}\nonumber\\
& &dv(t) = \theta(1-v(t))dt + \epsilon (t) \sqrt{v(t)}dZ(t)\nonumber\\
& &dZ(t) \cdot dW(t) = \rho dt,
\end{eqnarray}
where $dZ(t)$ is a Wiener process, $\lambda(t)$ and
$\gamma(t)$ are deterministic functions of time ($\gamma(t) \in
(0,1]$), $v(t)$ is a stochastic variable that drives the rate
variance, $\theta$ is a constant that estimates the mean reversion
speed of the process $v(t)$, $\rho$ is a constant that estimates
the correlation between the short rate and the volatility (hence
$\rho \in [-1,1]$),
and $\epsilon (t)$ is the volatility associated to $v(t)$, which in this case is simply a deterministic function of time.\\
Defining $x(t)=r(t)-f(0,t)$, we have
\begin{eqnarray}
dx(t) = [-\kappa x(t) + y(t)]dt + \eta (t,x(t)) dW(t),
\end{eqnarray}
where $dW\left(t\right)$ is another Wiener process.
It can be analytically shown that the price of the zero coupon bond in this formulation reads
\begin{eqnarray}
&&p(t,T) = \frac{p(0,T)}{p(0,t)}e^{-G(T-t)x(t)-\frac{1}{2}G(T-t)^{2}y(t)}\nonumber\\
&&G(s) = \frac{1-e^{ -\kappa s }}{\kappa} \label{gs},
\end{eqnarray}
The HJM model is thereby fully characterized. The state variables
belong to the following range:
\begin{eqnarray}
r(t)\in[0,\infty), \ v(t)\in[0,\infty), \ y(t)\in[0,\infty)
\nonumber
\end{eqnarray}

\subsection{Model reformulation in terms of Partial Differential
Equations} \label{model}

Given the above model formulation, now we would like to
price contracts whose future payoffs depend on the evolution of
the yield curve, that is to say, those payoffs are deterministic
functions of the ZCBP at certain times. Let $C(t)$ be the price in
a given time $t$, of a contract that pays in $T>t$ in terms of
$F(T,p(T,s)),s\geq T$, where \(P\) denotes the payoff function.
The simplest example of this situation is the zero coupon bond
whose payoff function is a constant function, \(F(T,p(T,s)) = 1\).
The so called Caplet~\cite{HWgeneral} is another example of a
payoff function, frequently traded in the market:
\begin{eqnarray}
F(T,p(T,T_{M})) = \max (1 - \Delta_M p(T,T_M), 0 ),
\end{eqnarray}
where $T$ is the contract's expiration date, $T_{M}$ is the
contract's payment date and \(\Delta_M = 1+ (T_M - T)K\), where
$K$ stands for the strike and is a positive constant. In this
paper we focus on both the Zero Coupon Bond and the Caplet as the derivatives under study.\\
In these terms, the price of an interest rate derivative depends
on time and space $C(t,x,y,v)$ and fulfills the following partial
differential equation:
\begin{eqnarray}
\frac{\partial C}{\partial t} + \zeta_{rr} \frac{\partial^{2}
C}{\partial r^{2}} + \zeta_{vv} \frac{\partial^{2} C}{\partial
v^{2}} + \zeta_{rv} \frac{\partial^{2} C}{\partial r \partial v} +
\mu_r \frac{\partial C}{\partial r}
 + \mu_v \frac{\partial C}{\partial v} + \mu_y \frac{\partial C}{\partial y} = r C \label{1},
\end{eqnarray}
where the Feynman-Kac formula \cite{Feynman} is applied to map the
stochastic problem into a PDE one (note that the financial
notation for the derivatives -the so called Greeks- is not used
here). The coefficients
\(\zeta_{rr},\zeta_{rv},\zeta_{vv},\mu_v,\mu_r,\mu_y\) are
functions of the space variables \(r,v,y\) and time, but their
dependence has been omitted for clarity. These coefficients are
given by

\begin{eqnarray}
& &\zeta_{rr} \equiv \frac{1}{2} \lambda(t) ^{2} r^{2\zeta(t)} v \nonumber\\
& &\zeta_{vv} \equiv \frac{1}{2} \epsilon(t)^{2} v \nonumber\\
& &\zeta_{rv} \equiv \lambda(t) r^{\zeta(t)} \epsilon(t) \rho v \nonumber \\
& &\mu_r \equiv \frac{\partial f(0,t)}{\partial t} - \kappa(r-f(0,t))+y \nonumber\\
& &\mu_v \equiv  \theta(1-v) \nonumber\\
& &\mu_y \equiv  \lambda(t) ^{2} r^{2\zeta(t)} v - 2\kappa y
\end{eqnarray}

The boundary/terminal conditions of the problem are different
depending on the payoff function.
For the cases of the zero coupon bond and the Caplet, these are given by:\\

\begin{itemize}
\item \emph{Zero coupon bond with expiration date $T$:} The
terminal condition simply reads
\begin{eqnarray}
C(T,r,y,v) = 1
\end{eqnarray}
In the limit $r\rightarrow\infty$ the payoff is zero, and
consequently one boundary condition is:
\begin{eqnarray}
C(t,r\rightarrow\infty,y,v) = 0,\ \ t<T
\end{eqnarray}
Now, when $r=0$ the PDE reduces to
\begin{eqnarray}
& &\frac{\partial C}{\partial t} + \left.\zeta_{vv}\right|_{r=0} \frac{\partial^{2} C}{\partial v^{2}} + \left.\mu_r\right|_{r=0} \frac{\partial C}{\partial r} + \left.\mu_v\right|_{r=0} \frac{\partial C}{\partial v} + \left.\mu_y\right|_{r=0} \frac{\partial C}{\partial y} = 0 \\
& &C(T,r\rightarrow0,y,v) = 1, \nonumber
\end{eqnarray}
In the boundary $y\rightarrow\infty$ the PDE reads
\begin{eqnarray}
& &\frac{\partial C}{\partial t} -  \kappa y \frac{\partial C}{\partial r}  -2\kappa y  \frac{\partial C}{\partial y} = 0 \\
& &C(T,r,y\rightarrow\infty,v) = 1 \nonumber
\end{eqnarray}
that is analytically solvable (Lagrange's method yields a solution
of the type $C=F(y e^{2kt}, 2r-y)$, where $F$ is a
generic function that must be determined to fulfill the boundary conditions). Finally, note that the boundaries $v\rightarrow\infty$ and $v\rightarrow 0$ are not relevant in this case as long as the zero coupon bond price is not a function of $v$.\\
\item \emph{Caplet}: The terminal condition is given by
\begin{eqnarray}
C(T,r,y,v) = \max (1- \Delta_M p(T,T_{M};r,y,v) ,0)
\end{eqnarray}
Just as in the case of the zero coupon PDE, the boundary condition
for $r\rightarrow\infty$ is
\begin{eqnarray}
C(t,r\rightarrow\infty,y,v) = 0,\ \ t<T.
\end{eqnarray}
When $r=0$ the PDE reduces to:
\begin{eqnarray}
& &\frac{\partial C}{\partial t} + \left.\zeta_{vv}\right|_{r=0}
\frac{\partial^{2} C}{\partial v^{2}}
+ \mu_r \frac{\partial C}{\partial r} +\left.\mu_v\right|_{r=0} \frac{\partial C}{\partial v} +\left.\mu_y\right|_{r=0}  \frac{\partial C}{\partial y} = 0 \\
& &C(T,0,y,v) = \max(1-\Delta_M p(T,T_{M};0,y,v),0) \nonumber
\end{eqnarray}
In the limit $y\rightarrow\infty$, the PDE and its boundary
condition have the following shape:
\begin{eqnarray}
& &\frac{\partial C}{\partial t} +  y \frac{\partial C}{\partial r}  - 2\kappa y  \frac{\partial C}{\partial y} = 0 \\
& &C(T,r,y\rightarrow\infty,v) = 0 \nonumber
\end{eqnarray}
Thereby we have:
\begin{eqnarray}
C(t,r,y\rightarrow\infty,v) = 0, t<T.
\end{eqnarray}
In the case $v\rightarrow\infty$ we will have
\begin{eqnarray}
C(T,r,y,v\rightarrow\infty) = p(t,T)-p(t,T_{M}),
\end{eqnarray}
and when $v=0$ the PDE reduces to
\begin{eqnarray}
& &\frac{\partial C}{\partial t} + \left.\mu_r\right|_{v=0} \frac{\partial C}{\partial r} +  \theta \frac{\partial C}{\partial v}  -2\kappa y \frac{\partial C}{\partial y} = r C \\
& &C(T,r,y,v\rightarrow0) = \max(1-\Delta_M p(T,T_{M};r,y,0),0)
\nonumber
\end{eqnarray}
that in this case can only be solved numerically.\\
Additionally, when $v=0$ the following identity is commonly
assumed to hold:
\begin{eqnarray}
\left.\frac{\partial^{2}C}{\partial
v^{2}}\right|_{v\rightarrow0}=0 \nonumber,
\end{eqnarray}
and will be taken into account in the numerical development.
\end{itemize}

\subsection{Model parameters}
\label{sec:ParametrosTIpicosDeMercadoParaElModeloHJM} In order to
fully specify the preceding PDE, the constants and functions
implicitly defined in the HJM model have to be initialized. As a
reference guide, and for the sake of an order of magnitude
estimation, we introduce the characteristic market parameters in 2007:\\

-\noindent \emph{Initial zero coupon curve}:
\begin{itemize}
\item $p(0,T) = 1.04^{-T} $
\end{itemize}

-\emph{Volatility function}:
\begin{itemize}
    \item $\kappa = 0.001$
    \item $\lambda(t) = 0.15$
    \item $\gamma(t) = 0.9$
    \item $\epsilon(t) = 1.5$
    \item $\theta = 0.25$
    \item $\rho = -0.75$
\end{itemize}
Observe that in a practical situation, for the calibration, values
of the parameters need to be optimized using \textit{e.g.} a
conjugated gradient or genetic algorithm, iterating several times
the model evaluation and comparing the results with the market
data. This is also supposed to be done in real time, and therefore
it is of fundamental importance that the model evaluation
(numerical method) is as efficient as possible. In this work we
focus on this fundamental issue and propose a numerical
methodology that enables an efficient evaluation of the model
(suitable for real time execution), while the global calibration
problem (the aforesaid optimization method) is not addressed.

\section{Optimizing the numerical scheme}
\label{numerical}

Before solving numerically equation (\ref{1}) and in order to optimize the numerical approach
and reduce its computational cost, we need to make some
preliminary analysis. Three points are of major
importance, namely (i) the study of some analytical solutions will
provide information about the system's solution itself, (ii) a
detailed scale analysis of the problem will help to optimize the mesh resolution, increasing it only where/when needed. Finally, some
considerations regarding the metric will also be addressed.\\

\subsection{Particular solutions}
The first case considers the solution of the zero coupon bond
curve with expiration date $T$. As far as this solution does not
depend explicitly on $v$ (non-arbitrage conditions yield a
volatility independent ZCBP), we assume $v=0$ without lack of
generality in order to simplify the system of equations.
Furthermore, constant $\kappa$ is generally small (volatility
function depends on $\kappa$) and we have thus assumed that it is
also null. Hence, the partial differential equation reduces to:
\begin{equation}
\frac{\partial C}{\partial t} + \left(\frac{\partial
f(0,t)}{\partial t} +y \right) \frac{\partial C}{\partial r} = r
C(t,r,y) \label{eqn1}
\end{equation}
with the additional condition $C(T,r,y)=1$. We can trivially map
this PDE into a system of ordinary differential equations of the
following shape:
\begin{equation}
\frac{dt}{d\tau}=1, \quad \frac{dr}{d\tau}= \frac{\partial
f(0,\tau)}{\partial \tau} +y , \quad  \frac{dC}{d\tau}= r C.
\label{eqn2}
\end{equation}
with initial conditions $\tau=0,\quad t=T,\quad r=r_1,\quad C=1$, and $y$ being a parameter.\\
It is indeed easy to check that its solution is:
\begin{equation}
C(t,r,y)= e^{-\int_t^T f(0,s)ds} e^{-(T-t)
(r-f(0,t))-\frac{(T-t)^2}{2} y}. \label{exacta}
\end{equation}
Note that not assuming a null value for $\kappa$ is equivalent to substitute $T-t$ by the function $G(T-t)$, which is defined in (\ref{gs}).\\
The usefulness of this analytical solution is twofold: first, it
will serve to validate the numerical method, and second, it will
stand as as boundary condition for the (more general) Caplet
problem.

\subsection{Scale analysis}
In the financial realm, reducing the computing time as well as the
computational cost (in terms of memory resource, for instance) is
fundamental. An adequate temporal and spatial scale analysis will
enable us to increase the mesh resolution only where/when needed,
what leads to a saving of computational resources. Typically, this
scale analysis is done by adimensionalizing the equations under
study and consequently comparing the relevance of the respective
terms (this technique is broadly used in fluid mechanics when
performing the scale analysis of the Navier-Stokes equations, for
instance~\cite{fluid}).\\
Scales are indeed determined by the variables characteristic
values, as well as by the boundary conditions. In the case of the
Cap problem, the variable of reference is the interest $r(t)$
($r\in[0,100]$, that is to say, a percentage). We can rescale this
variable as:
$$
\tilde r = r/r_0, \quad r_0 \simeq 10^{-2}
$$
in such a way that the characteristic value of $\tilde r$ is the
unity ($r_0$ is usually the forward rate of interest observed at
value date). Now, having in mind that the Cap's boundary condition
reads
$$C(T,0,y,v) =  \max(1-(1+(T_{M}-T) r_0 \tilde K)p(T,T_{M};0,y,v),0),$$
where  $\tilde K=K/r_0$, we may define the characteristic time
$\tilde t$ as:
$$
\tilde t= r_0 t.
$$
Finally, taking into account the relation between $r$ and $y$
(equation \ref{eqn2}):
$$
\Delta r \simeq y \Delta t, \rightarrow \tilde y= y/r_0^2.
$$
With these rescaled variables, the original equation takes the
following shape:
\begin{eqnarray}
r_0 \frac{\partial C}{\partial t} + h_1(t, r,v) \frac{\partial^{2}
C}{\partial r^{2}}
+ h_2(t,v) \frac{\partial^{2} C}{\partial v^{2}} + h_3(t,r,v) \frac{\partial^{2} C}{\partial r \partial v} +  \nonumber \\ \nonumber \\
+ h_4(t,r,v,y) \frac{\partial C}{\partial r}
 +  h_5(v) \frac{\partial C}{\partial v}
+  h_6(t,r,v,y) \frac{\partial C}{\partial y} = r r_0 C
\label{eqnfinal}
\end{eqnarray}
where
\begin{eqnarray}
h_1(t,r,v) = \frac{1}{2} \lambda(t) ^{2} r^{2\gamma(t)} r_0^{2(\gamma(t)-1)} v,  \quad h_2(t,v) = \frac{1}{2} \epsilon(t)^{2} v, \nonumber \\ \nonumber \\
h_3(t,r,v) = \lambda(t) r^{\gamma(t)} \epsilon(t)  \rho v r_0^{\gamma(t)-1}, \nonumber \\ \nonumber \\
h_4(t,r,v,y) = (r_0\frac{\partial f(0,t)}{\partial t} - \kappa(r-f(0,t))+ r_0 y), \nonumber \\ \nonumber \\
h_5(v) = \theta(1-v),  \quad h_6(t,r,v,y) = (\lambda(t) ^{2}
r^{2\gamma(t)} v -2\kappa y). \nonumber
\end{eqnarray}
(Note that the $\tilde{}$  marks have been eliminated for notation
simplicity). No rescaling has been applied to both $v$ and
$C(t,r,v,y)$ as long as there's no dominant scale defined (the
latter rescaling wouldn't affect the resultant equation).

\subsection{Metrics}\label{metrics}
By introducing metrics in the independent variables ($r,v,y$), we
can transform the problem's domain into a computational domain
which is usually simpler, and consequently concentrate the mesh
points in the areas under study. In our case of study (Cap),
provided that the solution is likely to live in the strike's
neighborhood (K), this should be the most dense zone. Following
Tavella \cite{Tavella},
 we have used an hyperbolic-like metric generally defined as:
$$
z = K + \alpha \sinh(c_2 x + c_1(1 - x)),
$$
$$
c_1= \mathrm{asinh} \left(\frac{(z_{0}-K)}{\alpha}\right), \quad
c_2=
 \mathrm{asinh}\left(\frac{(z_{\infty}-K)}{\alpha}\right),
$$

\begin{equation}
x \in [0,1], z \in [z_{0},z_{\infty}], \mbox{  being   }  z=(r,v,
\mbox{ or } y). \label{metrica}
\end{equation}
In order to get a mesh that is accurate enough in the zone under
study, the parameters $\alpha, K, z_0, z_{\infty} $
have to be correctly chosen for each dimension of the problem.\\
It is worth saying that the introduction of these kind of metrics
doesn't modify the system of equations in a substantial manner.
Note that the derivatives with respect of the new variables can be
expressed as:

$$
\frac{\partial C}{\partial r} = \frac{\partial C}{\partial x}
\frac{1}{\partial r/\partial x}
$$
and
$$
\frac{\partial^2 C}{\partial r^2} = \frac{\partial^2 C}{\partial
x^2} \frac{1}{\left(\partial r/\partial x\right)^2}-
 \frac{\partial C}{\partial x} \frac{\partial ^2 r/\partial x^2}{\left(\partial r/\partial x\right)^3}
$$
Thereby, if we denote $ J_z=\partial z/\partial x$, and $
J_{2z}=\partial z^2/\partial x^2$, where $z$ is a generic variable
from $(r,v,y)$ and $x$ its respective transform, the original
system of equations (\ref{eqnfinal}) will only differ from the new
one in the substitution of functions $h_i$, $i=1..5$ by:
$$
g_1(t,x_r,x_v)=h_1(t,x_r,x_v)/J_r^2, \quad
g_2(t,x_v)=h_2(t,x_v)/J_v^2,
$$
$$
 g_3(t,x_r,x_v)=h_3(t,x_r,x_v)/J_r/J_v, \quad g_4(t,x_r,x_v,x_y)=h_4(t,x_r,x_v,x_y)/J_r-h_1(t,x_r,x_v) J_{2r}/J_r^3,
 $$
 $$
 g_5(t,x_v)=h_5(x_v)/J_r-h_2(t,x_v) J_{2v}/J_v^3, \quad g_6(t,x_r,x_v,x_y)=h_6(t,x_r,x_v,x_y)/J_y.
$$
While hitherto we have only defined metrics in the direction of
the independent coordinates, it is actually possible to employ
more complex transformations that involve several variables
($\xi=\xi(r,v,y)$). However, it is likely that the integration
domain wouldn't in that case be cartesian anymore, and
consequently the finite difference scheme wouldn't apply.
Furthermore, while these types of metrics would eventually enable
us to eliminate the cross derivative terms in (\ref{eqnfinal})
(transforming the original equation into its canonical form), this
transformation would on the other side modify the frontiers of the
problem from straight to curve lines, something
that not desirable in any case.\\
Once the preliminary insights have been put forward, we will
describe in the next sections the numerical method employed to
integrate equation (\ref{eqnfinal}) as well as the results that we
have obtained.

\section{Numerical methods}
\label{nummethod}

As commented above, the equation under hands is (\ref{eqnfinal}):
\begin{eqnarray}
r_0 \frac{\partial C}{\partial t} + g_1(t, r,v) \frac{\partial^{2}
C}{\partial r^{2}}
+ g_2(t,v) \frac{\partial^{2} C}{\partial v^{2}} + g_3(t,r,v) \frac{\partial^{2} C}{\partial r \partial v} +  \nonumber \\
+ g_4(t,r,v,y) \frac{\partial C}{\partial r}
 +  g_5(v) \frac{\partial C}{\partial v}
+  g_6(t,r,v,y) \frac{\partial C}{\partial y} = r r_0 C.
\end{eqnarray}
Let us define a new temporal variable
$$
t= T-t
$$
where $T$ stands for the maturity. The integration should then be
done for $t\in [0,T]$.
Note that equation (\ref{eqnfinal}) is parabolic for $r$ and $v$ and hyperbolic for $y$. \\

In the particular problems concerning the estimation of financial
derivatives, the execution time of numerical tools is an issue of
fundamental importance. According to this fact, it seems suitable
to apply second order schemes for the discretization of both
temporal and spatial partial derivatives, as far as these schemes
show optimal computational cost and adequate precision. In a
second step, one has to decide whether to apply explicit or
implicit schemes. Naturally, the simplest option is always to
tackle explicit schemes, which are fast and easy to implement.
However, it is easy to check that due to the second derivatives,
the following relation holds for the temporal and spatial
resolutions

$$\Delta t \simeq (\Delta r)^2.$$

This relation implies that in order to achieve a precision of say
$10^{-4}$ in the solution, a time step would need $10^{4}$
iterations. Moreover, the coefficients of the derivatives are
powers of $r$, $v$ or $y$, and the integration domain ranges to
the infinite. Since the time step is inversely proportional to
those coefficients, the problem comes to be even more delicate. We
can thus conclude that explicit temporal schemes won't fit in this
case due to their inevitably lengthy behavior. Thereby, we will
have to choose implicit schemes for the temporal integration.
These have the following general expression:

$$\frac{dC}{dt} = F(C) \rightarrow C^{n+1}-C^{n} = \Delta t ~( \theta F(C^{n+1}) + (1-\theta)
F(C^{n})),$$ where we have employed the usual numerical methods
notation $C^n\equiv C(t^n)$ and $t^n=n\Delta t$. Here $\theta$
stands for an explicit scheme for $\theta =0$ (Euler scheme) while
it stands for implicit schemes when $\theta \neq 0$. More
concretely, $\theta =1$ characterizes the so called Euler implicit
scheme and finally $\theta =1/2$ characterizes the second order
Crank-Nicholson scheme. We will use the latter one as the temporal
integrator as it is adequate to be used within ADI
schemes (this will be explained further in the text).\\
The first, second and cross spatial derivatives, are discretized
by centered finite difference schemes as it follows:

\begin{eqnarray}
 \delta_{xx}u\equiv\frac{\partial^2 u}{\partial x^2} = \frac{u_{i+1}-2 u_i + u_{i-1}}{\Delta x^2}, \quad
 \delta_x u\equiv\frac{\partial u}{\partial x} = \frac{u_{i+1}- u_{i-1}}{\Delta
 x},\nonumber \\
 \delta_{xz}u\equiv\frac{\partial^2 u}{\partial x \partial z} = \frac{u_{i+1,j+1}- u_{i+1,j-1} -u_{i-1,j+1}+ u_{i-1,j-1}}{\Delta x \Delta z}
 \label{esquema}
 \end{eqnarray}

  where $x,z$ is a generic variable that stands for $r$, $v$ or $y$. The notation $\delta_{xx} u$, $\delta_x
  u$ and $\delta_{xz} u $  describes the above difference schemes.
  Gathering both spatial and temporal schemes, we come to a final discretization of the
  following kind:
\begin{equation}
U^{n+1}-U^{n} = \Delta t  ( \theta F(U^{n+1}) + (1-\theta)
F(U^{n})) \label{sol1}
\end{equation}
where
\begin{eqnarray}
F(U) = g_1(t, r,v) \delta_{rr} U
+ g_2(t,v) \delta_{vv} U  + g_3(t,r,v) \delta_{rv} U \nonumber  \\
+ g_4(t,r,v,y) \delta_{r} U +  g_5(v) \delta_{v} U
+  g_6(t,r,v,y) \delta_{y} U  - r r_0 U, \nonumber \\ \nonumber\\
 \mbox{ and } U=\{C_{ijk}, i=0..nr, j=0..nv, k=0..ny\},
 \label{discreta}
 \end{eqnarray}
is the discretized solution vector in a structured mesh of dimension ($nr,nv,ny$).\\
It is worth saying that the use of centered difference schemes
allows us to obtain a compact stencil. For instance,
 note that the discretized equation in the point $C_{ijk}$ only contains information of $\{C_{i+1,j,k}, C_{ijk}, C_{i-1jk}\}$. Focusing on variable
$r$, equation ($\ref{sol1}$) would adopt the shape:
$$
U^{n+1}-U^{n} = \Delta t (( \theta L_r U^{n+1} +b ) + (1-\theta)
(L_r U^{n} + b)),
$$\\
where $L_r$ is a tridiagonal matrix \cite{libro general}
representing the spatial discretization of $g_1(t, r,v)
\delta_{rr}+g_4(t,r,v,y) \delta_{r}$ in $F(U)$ (equation
(\ref{discreta})), according to the discretization schemes
depicted in equation (\ref{esquema}). $L_v$, $L_y$ and $L_{rv}$
will be defined equivalently (see below).\\
Tridiagonal systems are indeed quite easy to implement and solve
(for instance, the Thomas algorithm~\cite{numerical} solves a
tridiagonal system in $7N$ operations, where N is the order of the
system). This goodness
will enable the use of Alternating Direction Implicit schemes (ADI) \cite{ADIseminal1, libro general} as will be shown further in the text.\\
Finally, taking into account that the equation is indeed linear,
it can be written as:
\begin{eqnarray}
U^{n+1}-U^{n} = \theta  \Delta t \left[L_r(t^n,r_i,v_j,y_k) U^{n+1} + L_v(t^n,v_j) U^{n+1}+  L_y(t^n,r_i,v_j,y_k) U^{n+1}\right]  \nonumber  \\
+ (1-\theta)  \Delta t \left [L_r(t^n,r_i,v_j,y_k) U^{n} +
L_v(t^n,v_j) U^{n}+
  L_y(t^n,r_i,v_j,y_k) U^{n}\right] \nonumber \\
 +   \Delta t L_{rv}(t^n,r_i,v_j) U^{n} +  \Delta t B(t^n,r_i,v_j), \nonumber
\end{eqnarray}
and realigning,
\begin{eqnarray}
\left[I- \theta \Delta t (L_r(t^{n+1/2},r_i,v_j,y_k)+ L_v(t^{n+1/2},v_j)+  L_y(t^{n+1/2},r_i,v_j,y_k)\right]( U^{n+1}-U^{n})  \nonumber  \\
=  \Delta t \left[L_r(t^{n},r_i,v_j,y_k) + L_v(t^{n},v_j)+  L_y(t^{n},r_i,v_j,y_k) + L_{rv}(t^n,r_i,v_j)\right] {U^n}+  \Delta t B(t^n,r_i,v_j) \nonumber \\
=  \Delta t F(U^n ,t^{n},r_i,v_j,y_k). \label{adiecu}
\end{eqnarray}
Note that the operators $L_r, L_v, L_y $ include the terms related
to the spatial discretization. These operators, treated
implicitly, give rise to a system of equations($I- \theta
(L_r+L_v+L_y))U=f$) which
 in general has 7 diagonals, and whose resolution can be performed applying either direct or iterative methods.
  However, each one of them treated separately can be rewritten as a tridiagonal matrix, whose resolution is trivial as commented
  above. Note also that the operators $L_r, L_v, L_y $ have Neumann boundary conditions
and consequently do not include any Dirichlet-like information.\\

 Finally, note that the mixed derivative term ($L_{rv}$) is only treated explicitly, because
otherwise its inclusion in the implicit scheme would eliminate the
tridiagonal structures, and would consequently avoid the use of
ADI schemes that will be described in what follows. This fact does
not affect in any case neither to the convergence nor the
precision of the numerical solution. Detailed numerical analysis and validation of the mixed derivative term has been
already performed by different authors: examples of implementation for the fluid mechanics problems
 are given in \cite{McKee,goetz}, and a detailed analysis of  numerical stability analysis
and convergence can be found in \cite{newADI}.

\subsection{Alternating Direction Implicit (ADI) schemes}
\label{adi}

The ADI schemes belong to the category of Splitting methods
\cite{ADIseminal1, ADIseminal2, libro general, Steger}, used in the resolution of multidimensional PDE systems.
The key idea behind these methods is to separate the original
multidimensional problem in several unidimensional split problems.
Then, each split problem can be under certain conditions reduced
to the resolution of a tridiagonal system of equations. These
conditions are related to the use of centered spatial
operators in structured meshes, which is our case.\\
ADI schemes were initially introduced by Douglas, Peaceman and
Rachford \cite{ADIseminal1, ADIseminal2} in order to integrate,
using finite difference schemes, the well known Navier-Stokes
equations describing the fluid motion. Some modifications have
been put forward so far (see for instance \cite{Steger,McKee,
newADI, libro general}, in order to apply these schemes to either
stationary or non stationary problems. In this work we will use an
ADI scheme recently put forward by Hout \& Welfert \cite{newADI},
called the Douglas scheme.\\

Consider equation (\ref{adiecu}), this one can be formally written
as:
$$
[I-\theta  \Delta t ( L_r+L_v+L_y) ]\Delta U^n = \Delta t F(U^n)
$$

The Douglas scheme applies thus in the following way:

$$
\Delta U^{0}=  \Delta t F(U^n)
$$
$$
[I-\Delta t \ \ Lr \ \ \Delta \tilde U^1]=\Delta U^0
$$
$$
[I-\Delta t  \ \ Lv \ \ \Delta \tilde U^2]=\Delta \tilde U^1
$$
$$
[I-\Delta t  \ \ Ly  \ \ \Delta \tilde U^n]=\Delta \tilde U^2
$$
$$
U^{n+1}=U^n+\Delta \tilde U^n
$$

Note that each step only requires the resolution of a tridiagonal
system of dimension $nr$, $nv$, or $ny$. The unconditional
convergence of this scheme has been proved for $\theta = 0.5$
(Crank-Nicholson) in 2-dimensional systems with constant
coefficients\cite{libro general}. However no similar study has
been performed so far in the 3-dimensional case with variable
coefficients \cite{newADI}, which is nonetheless our case.
Special attention will be thus paid to the convergence behavior of the solution.\\

It is worth saying at this point that the Craig \& Sneyd \cite{craig} method is
an apparent improvement to the Douglas scheme (in terms of the
solution precision) when mixed derivative are present, while being more expensive computationally
speaking. We actually have also tackled this ADI scheme, but given
that no such improvement has been observed, we will only focus on
the Douglas scheme.

Finally, the Douglas scheme, as any other ADI scheme, is an Approximate Factorization (AF) of the original equations
with the errors of order $\Delta t^3$ for 3D problems. An efficient subiteration procedure can be applied to eliminate  the AF.
This method, known as Huang's approximate factorization correction \cite{goetz}, has been also checked in this context, but it is
computationally more  expensive and no additional improvements have been observed in the range of accuracy we are working.

\section{Validation}
\label{validation}

We have done two kind of studies in order to validate the numerical methods:\\
(i) first, we have compared the numerical solution of the zero
coupon curve or deduction curve with its analytical solution, in
two different situations, and\\
(ii) second, we have compared the numerical solution of a Caplet
with the one obtained by a 2D-Heston model \cite{libro general}
(it is easy to check that the model under study behaves, for
$T_M\rightarrow T$, as a Heston model for the libor rate with
identical parameters).

\subsection{Zero coupon curves} 
\begin{figure}[htpb]
\begin{center}
\includegraphics[width=8cm,height=8cm]{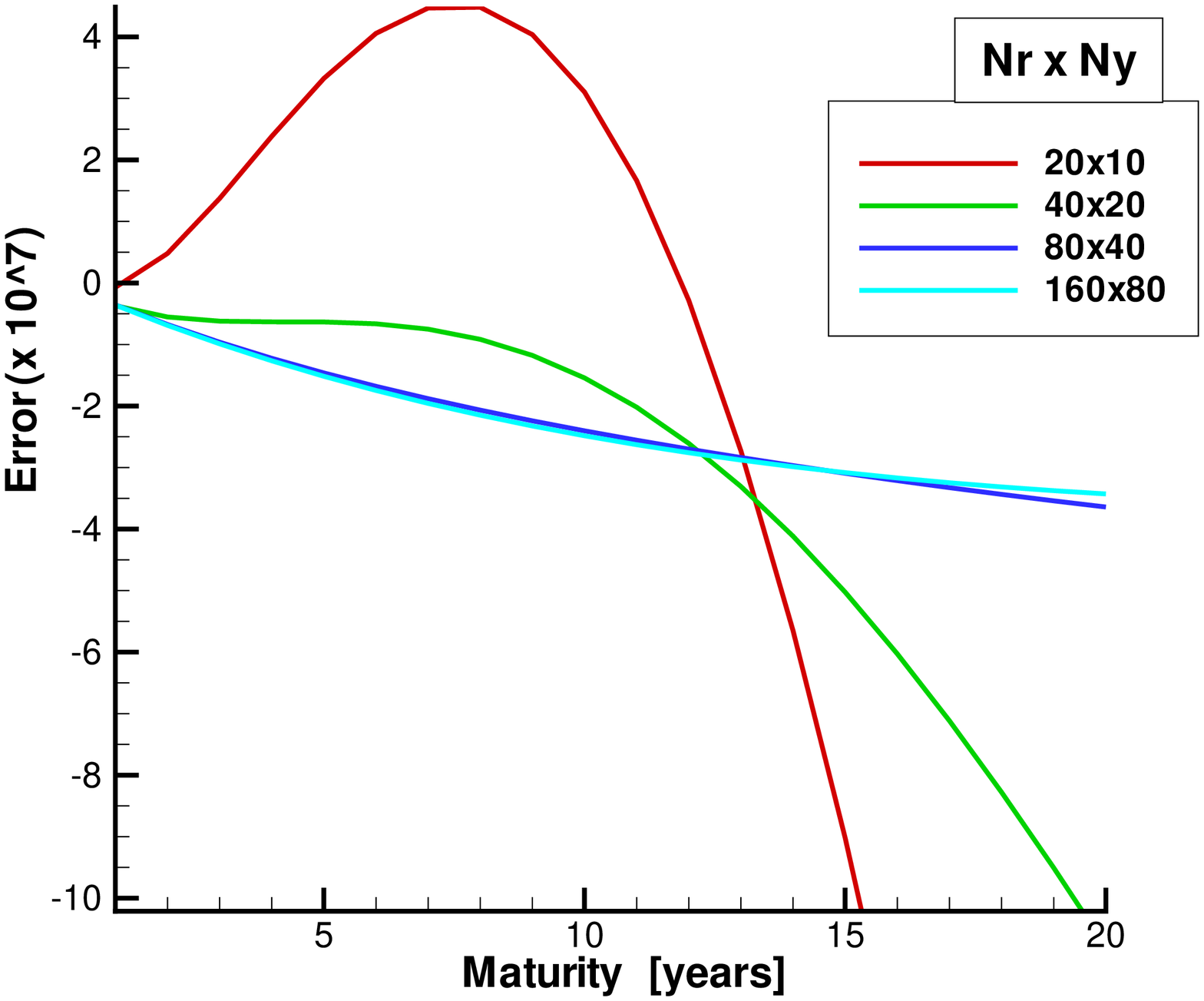}
\caption{Plot of the numerical solution's error as a function of
the mesh size, for the initial zero coupon curve.
 The particular values of the metric are $\alpha_r=0.05, \alpha_y=0.5, r_{\infty}=25, y_{\infty}=250.$}
 \label{fig1}
\end{center}
\end{figure}

\begin{figure}[htpb]
\begin{center}
\includegraphics[width=8cm,height=8cm]{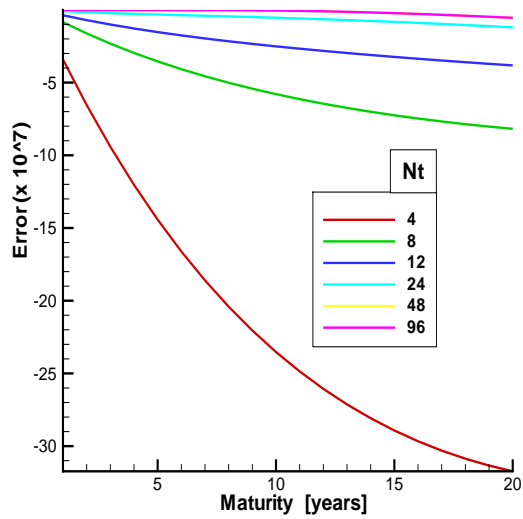}
\caption{Plot of the numerical solution's error (convergence) as a
function of the number of time steps per year, for the reference
mesh of size 100x40. The metric parameters are the same as for
figure 1.}
\label{fig2}
\end{center}
\end{figure}

In the first study we consider
the parameters and the initial zero coupon curve depicted in
section \ref{sec:ParametrosTIpicosDeMercadoParaElModeloHJM}.
Thereby, the forward rate of interest is constant
$f(0,t)=\log(1.04)$ and it has a null derivative. In figures
(\ref{fig1}-\ref{fig2}) we compare the theoretical zero coupon
curve with the one obtained through numerical simulations with
$r=\log(1.04)$, $y=0$, and $t=T$. Concretely, figure (\ref{fig1})
shows the error for different meshes. Note that this one is always
below $10^{-5}$ even for coarse meshes. As a result, we have set
the mesh reference values to $100\times40$. In figure (\ref{fig2})
we plot the convergence of the solution as a function of the
number of time steps per year. Notice that from 12 steps per year,
in a given mesh the variations are quite small ($O(10^{-7})$).\\

\begin{figure}[htpb]
\begin{center}
\includegraphics[width=8cm,height=8cm]{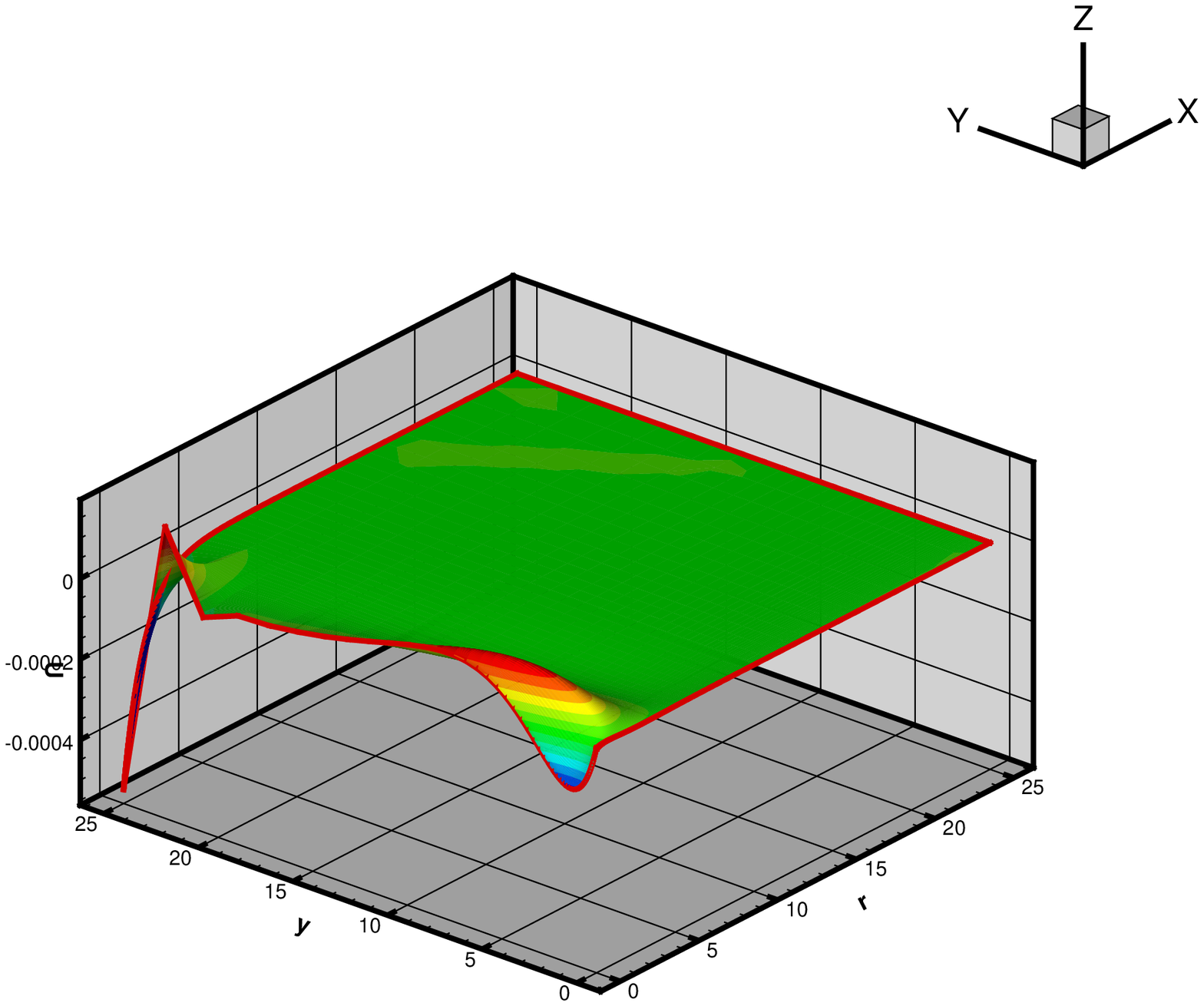}
\caption{Error spatial distribution obtained in t=T=20 years for
the zero coupon curve. The metric parameters are $\alpha_r=0.05,
\alpha_y=0.5, r_{\infty}=25, y_{\infty}=25.$, the mesh is the
reference one and the temporal discretization assumes 12 time
steps per year.}
\label{fig3}
\end{center}
\end{figure}

\begin{figure}[htpb]
\begin{center}
\includegraphics[width=8cm,height=8cm]{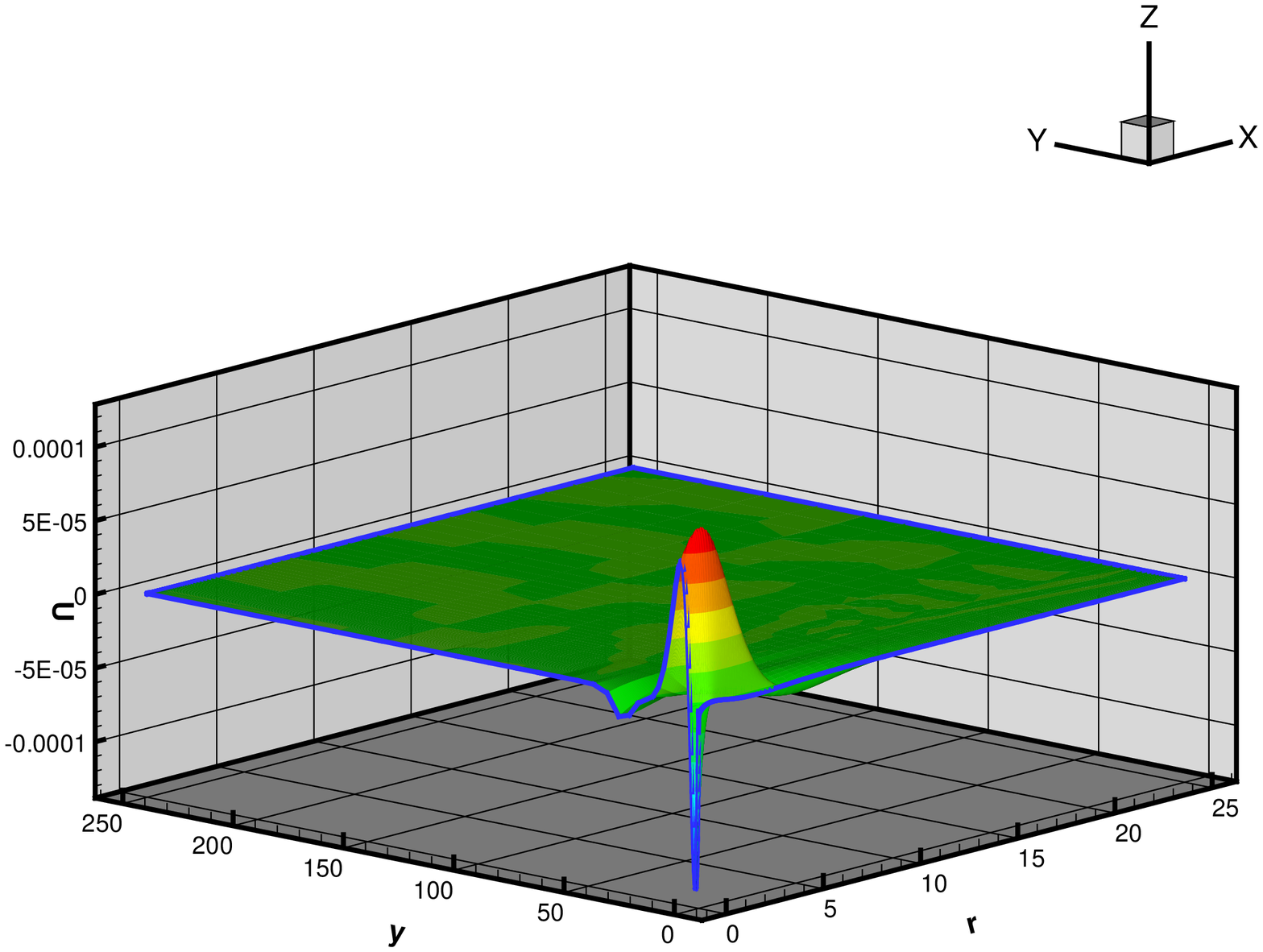}
\caption{Error spatial distribution obtained in t=T=20 years for
the zero coupon curve. The metric parameters are $\alpha_r=0.05,
\alpha_y=0.5, r_{\infty}=25, y_{\infty}=250.$, the mesh is the
reference one and the temporal discretization assumes 12 time
steps per year.}
\label{fig4}
\end{center}
\end{figure}

\noindent In figures (\ref{fig3}) and (\ref{fig4}) we show the error's
spatial distribution for a given set of parameters,
assuming $y_{\infty}=25$ (figure (\ref{fig3})) or $y_{\infty}=250$
(figure (\ref{fig4})) respectively. Note that in the former case, some non
desirable errors take place in the infinite boundary, which can
actually propagate into the zone under study $y\simeq 0$ (figure
(\ref{fig3})).

\begin{figure}[htpb]
\begin{center}
 \includegraphics[width=8cm,height=8cm]{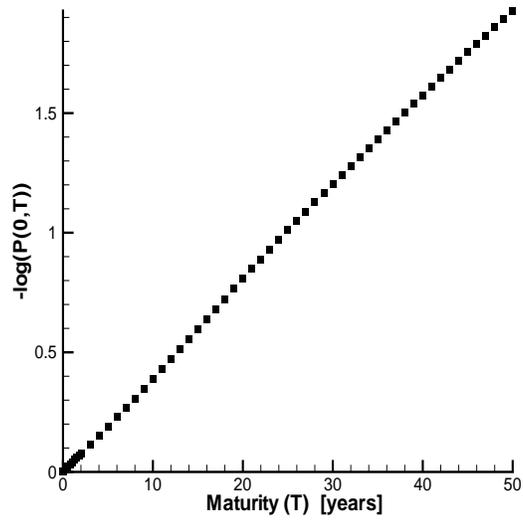}
 \caption{Initial zero coupon curve (discrete curve). Note that the plot is in semilog.}
 \label{fig5}
\end{center}
\end{figure}

\begin{figure}[htpb]
\begin{center}
 \includegraphics[width=8cm,height=8cm]{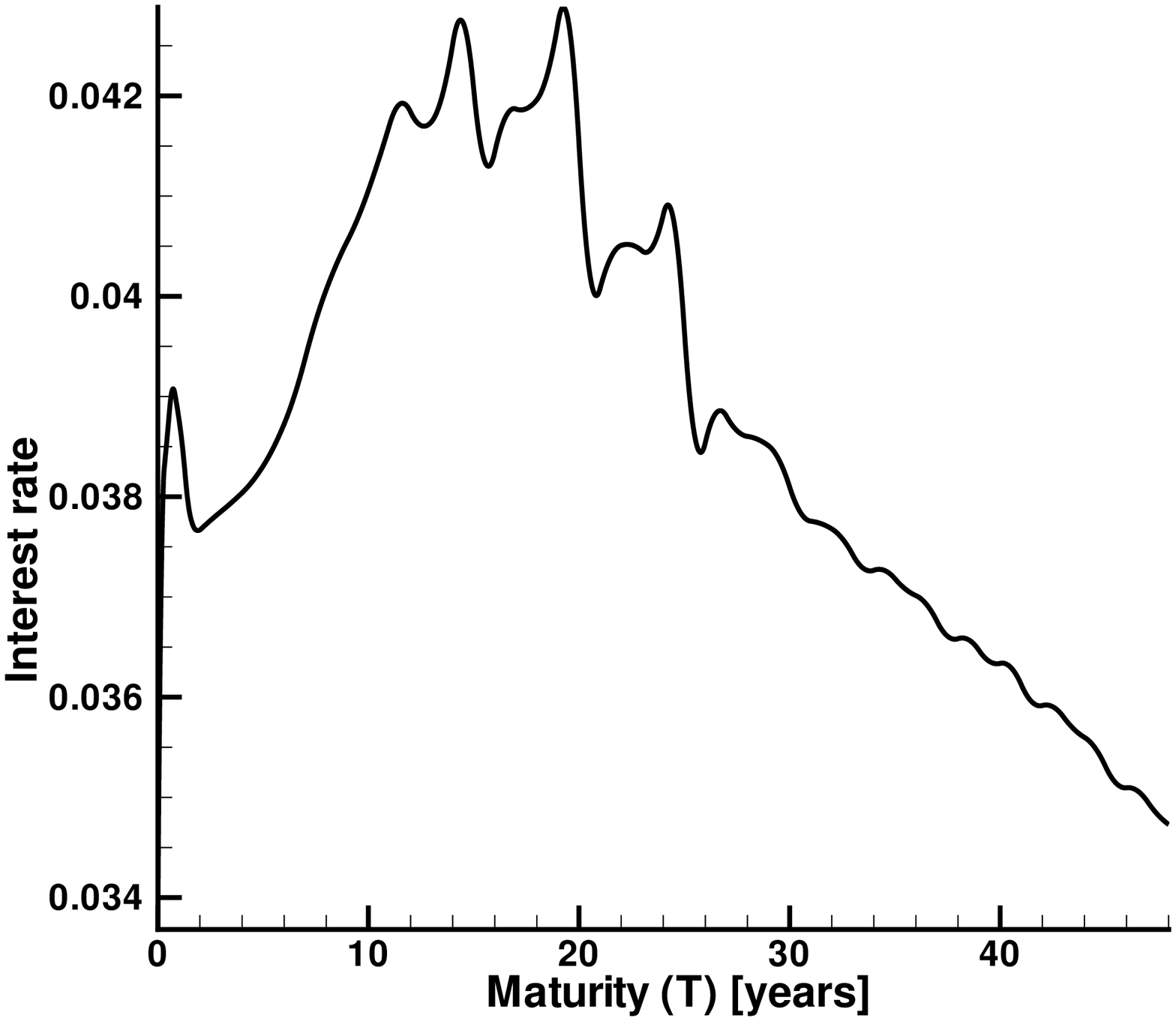}
\caption{Approximation of the forward rate of interest curve,
obtained with splines.}
\label{fig6}
\end{center}
\end{figure}

\begin{figure}[htpb]
\begin{center}
  \includegraphics[width=8cm,height=8cm]{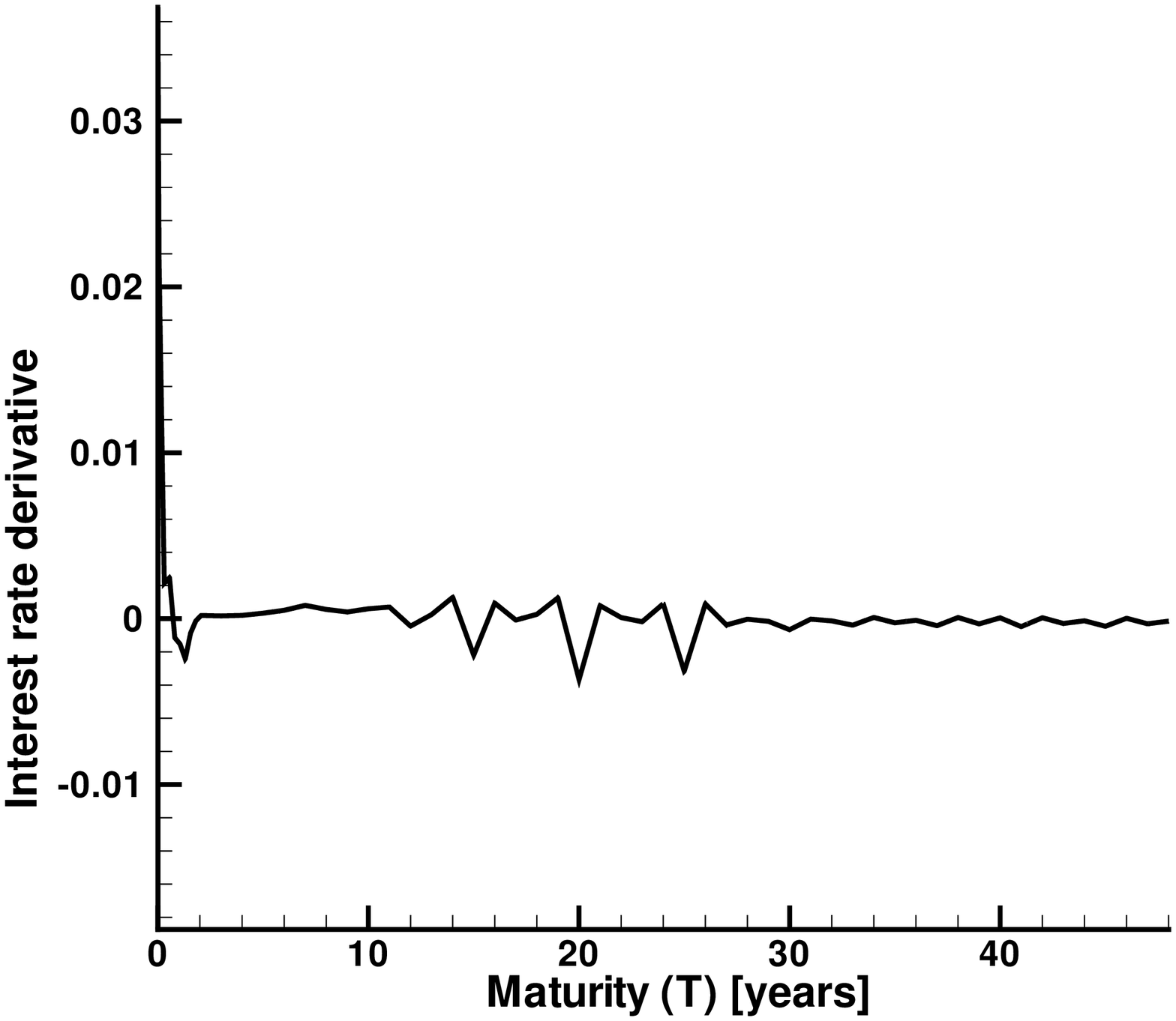}
\caption{Approximation of the  forward rate of interest curve
derivative, obtained with splines.}
\label{fig7}
\end{center}
\end{figure}

Up to know the numerical method is validated, as far as the
solution's error is confined, in the zone under study, around
$10^{-7}$. However, as long as the analytical solution strongly
depends on the initial curve, it is necessary to check whether if
the precision of the numerical method holds for more realistic
curves (with non null forward rate interest curve derivative). For
that task, in a second example we tackle a new initial zero coupon
curve, which is not anymore a continuous curve but a discrete
valued one (figure (\ref{fig5})). Its derivative is plotted in
figure (\ref{fig6}) and stands for the forward rate of interest
curve, and its second
derivative is plotted in figure (\ref{fig7}).\\

As long as the curve is expressed in terms of discrete values, we
need to perform a smoothing approximation in order to introduce it
in the simulation. Notice that the solution's smoothness will
strongly depend with the smoothness of this initial curve (this is
due to the fact that the temporal derivative of the solution is
related to the derivative of the forward rate of interest curve).
An appropriate solution to this problem is to approximate the
initial curve with splines~\cite{numerical}, in order to have a
piecewise function with continuous second derivative $df(0,T)/dT$,
and consequently have a solution with continuous temporal
derivative.

\begin{figure}[htpb]
\begin{center}
\includegraphics[width=8cm,height=8cm]{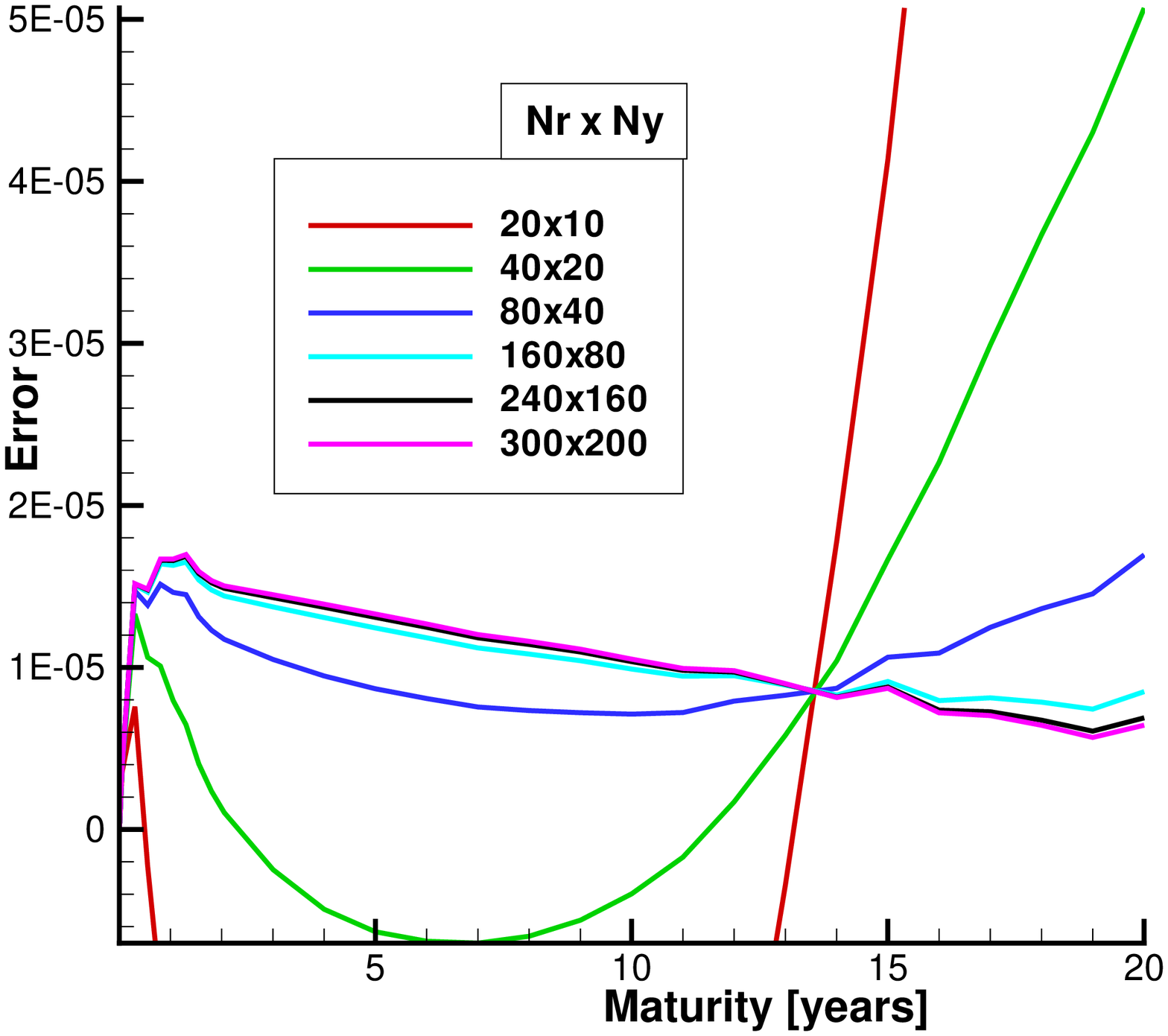}
\caption{Plot of the numerical solution's error as a function of
the mesh size $N_r \times N_y$, for the initial zero coupon curve.
 The particular values of the metric are $\alpha_r=0.05, \alpha_y=0.5, r_{\infty}=25, y_{\infty}=250.$}
 \label{fig8}
\end{center}
\end{figure}

\begin{figure}[htpb]
\begin{center}
 \includegraphics[width=8cm,height=8cm]{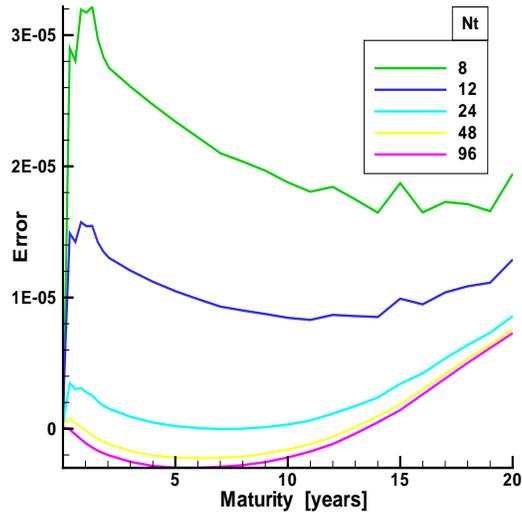}
\caption{Plot of the numerical solution's error (convergence) as a
function of the number of time steps per year, for the reference
mesh of size 100x40. }
\label{fig9}
\end{center}
\end{figure}

\begin{figure}[htpb]
\begin{center}
\includegraphics[width=10cm,height=10cm]{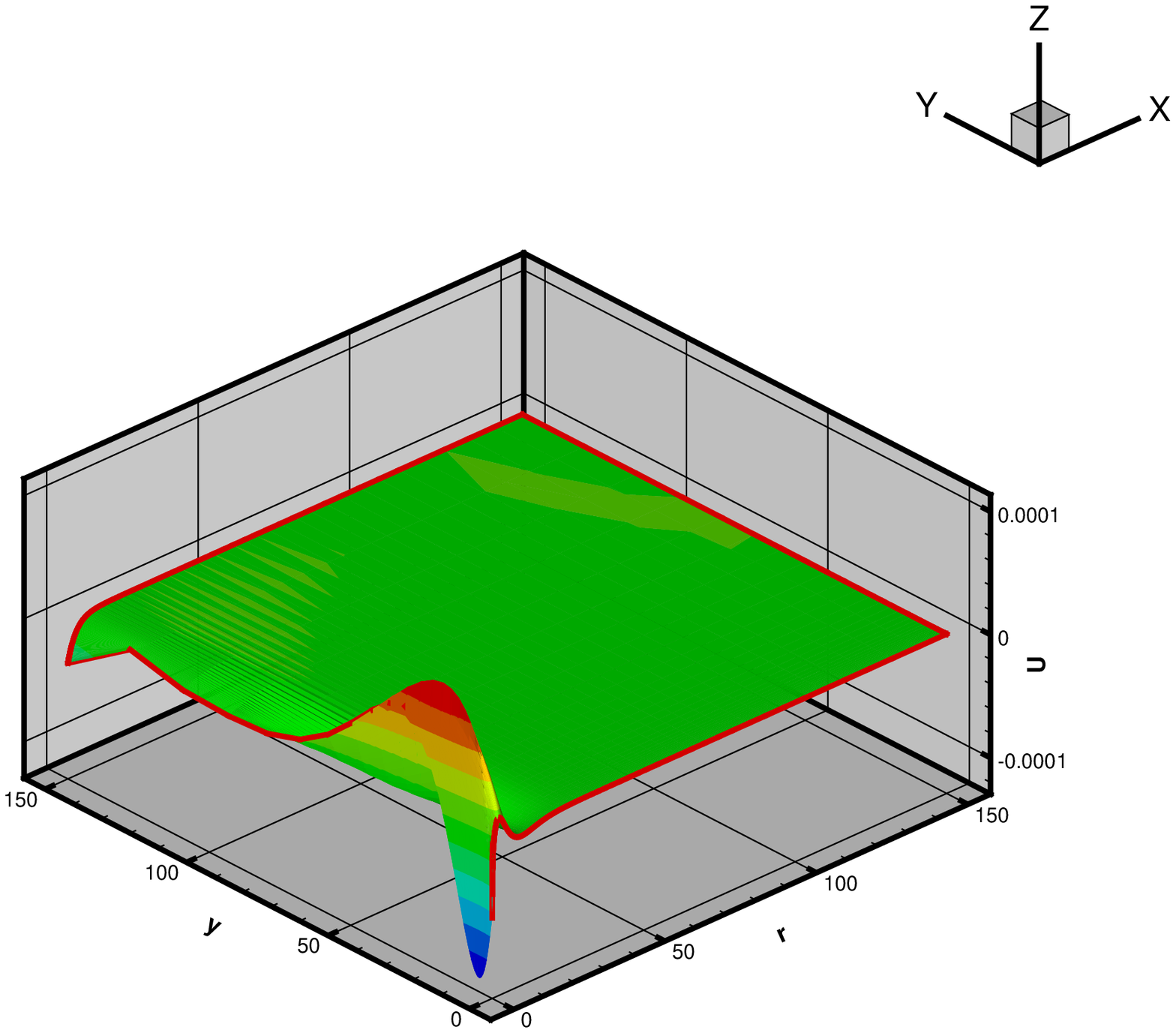}
 \caption{Error spatial distribution obtained in t=T=20 years for the zero coupon curve.
The metric parameters are $\alpha_r=0.05, \alpha_y=0.5,
r_{\infty}=150, y_{\infty}=150.$, the mesh is the reference one
and the temporal discretization assumes 12 time steps per year.}
\label{fig10}
\end{center}
\end{figure}

According to this approximation, we have performed the same
simulations and analysis as for the first study. Conclusions are
plotted in figures (\ref{fig8}-\ref{fig10}). As expected, the fact
that the derivative of the forward rate of interest curve is non
null has a net effect in the precision of the solution. While it
is quite easy to achieve convergence of order O($10^{-5}$), it
comes necessary to overrefine the mesh (figure (\ref{fig8})) or
alternatively increase the number of time steps (figure
(\ref{fig9})) in order to go beyond $10^{-6}$. Nevertheless, as is
shown in figure (\ref{fig10}), the error's spatial
 distribution is quite similar to the one found in the first
 study: we can conclude that the numerical method correctly
 reproduces the expected results.

\subsection{Caplet}

\begin{figure}[htpb]
\begin{center}
 \includegraphics[width=10cm,height=10cm]{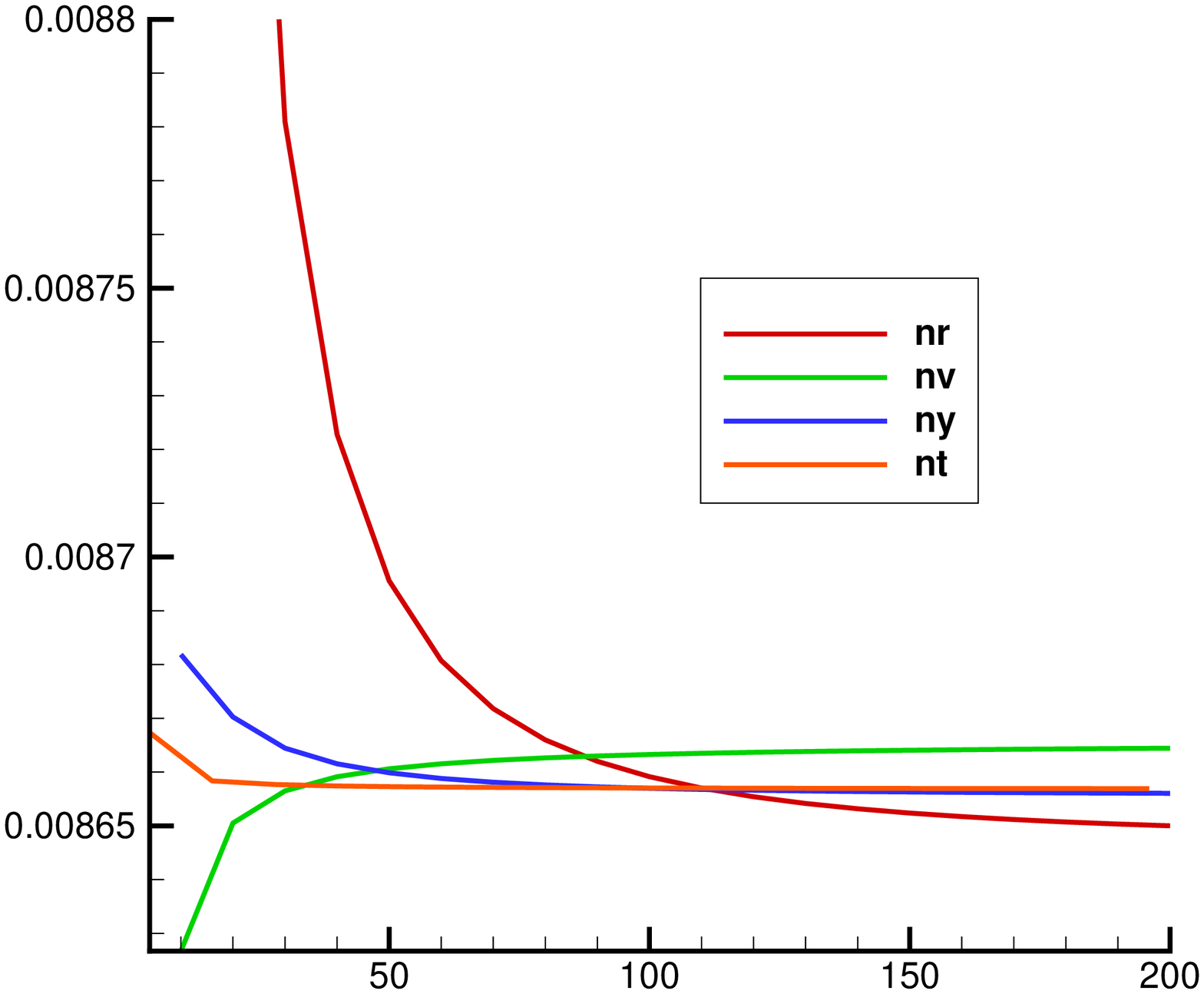}
 \caption{Premium variation for a generic Caplet as a function of nr, nv, ny, nt.}
 \label{fig11}
\end{center}
\end{figure}

The second validation test consists in making a comparison between
the results obtained with several Caplets and those obtained by
the Heston model, which is an already validated model \cite{libro
general}.\\
In order to optimize the mesh's size, we have performed a previous
analysis of the numerical scheme's convergence (both in spatial
and temporal discretizations). Some of the results are plotted in
figure (\ref{fig11}), where we represent the evolution of a
generic Caplet's prime  as a function of $nr$, $nv$, $ny$ and
$nt$. Notice that we need at least a mesh size of 100x40x40 if we
seek variations of the prime below $10^{-5}$.
With a reference mesh of 100x40x40, the number of time steps does not affect practically the results.\\

\begin{figure}[htpb]
\begin{center}
\includegraphics[width=8cm,height=8cm]{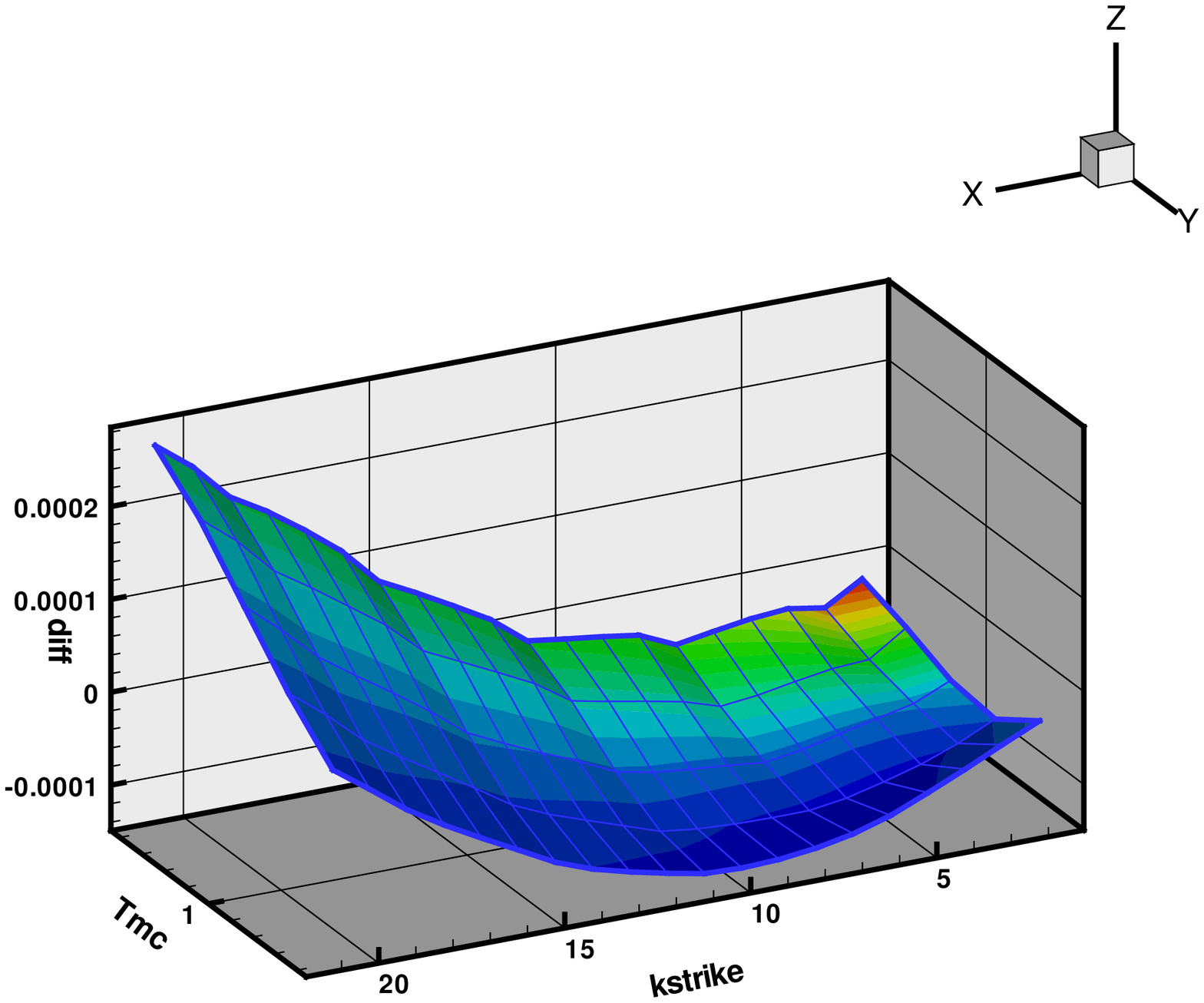}
\caption{Difference between the Heston model and the HJM model in
the assessment (premium) of several Caplets. Metric parameters:
$\alpha_r=0.5, \alpha_v=0.5, \alpha_y=0.5, r_{\infty}=250,
v_{\infty}=30, y_{\infty}=250.$ and 12 time steps per year.}
\label{fig12}
\end{center}
\end{figure}

\begin{figure}[htpb]
\begin{center}
  \includegraphics[width=8cm,height=8cm]{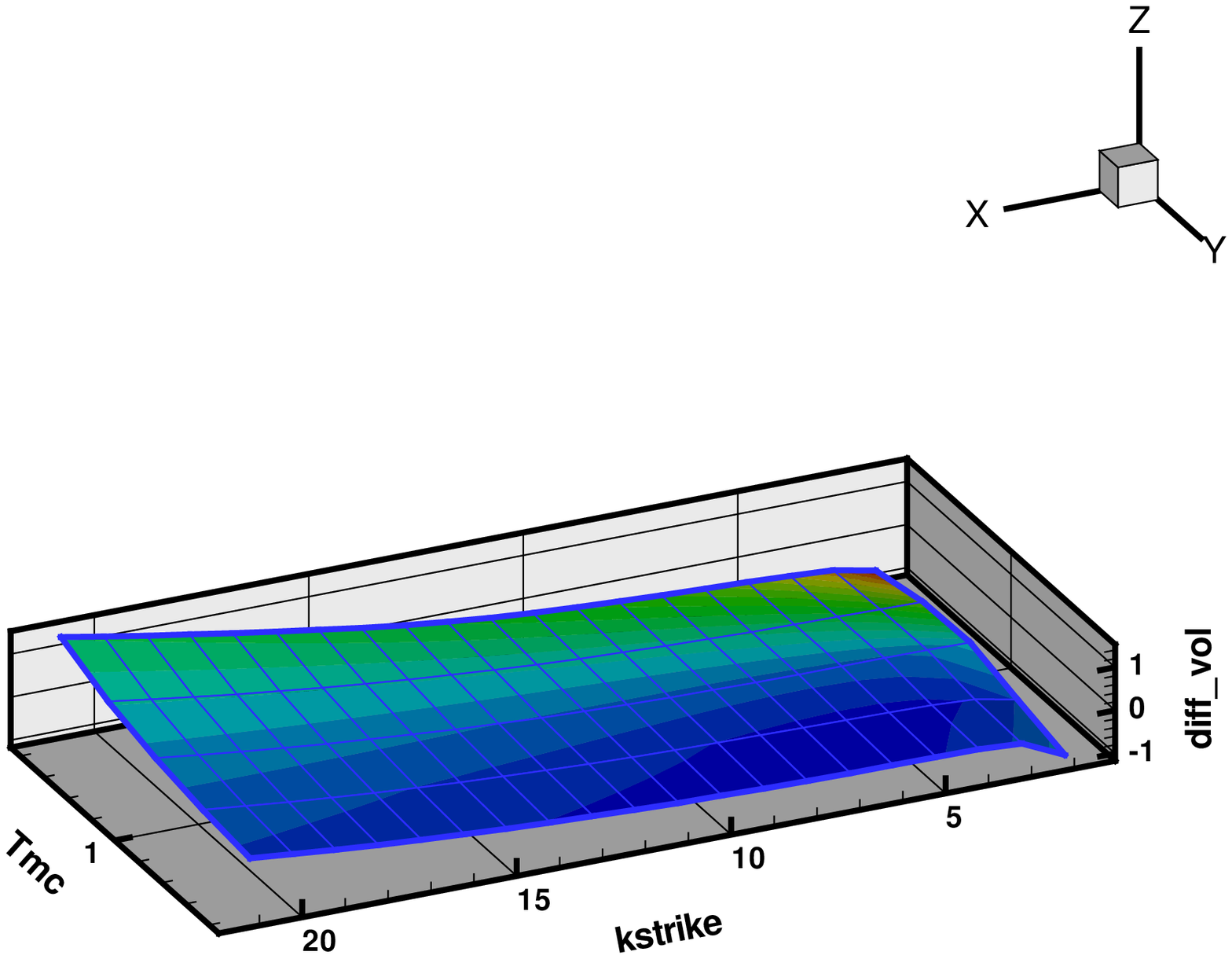}
  \caption{Difference between the Heston model and the HJM model in the assessment (volatility) of several Caplets.
Metric parameters: $\alpha_r=0.5, \alpha_v=0.5, \alpha_y=0.5,
r_{\infty}=250, v_{\infty}=30, y_{\infty}=250.$ and 12 time steps
per year.}
\label{fig13}
\end{center}
\end{figure}

The difference between the solutions of the two models are plotted
in figures (\ref{fig12}) and (\ref{fig13}), both for the premium
and for the volatility.

\subsection{Computation time}
In the following table we
have plotted, as a reference guide, the required computation
time for different Caplets. Simulations have been run in a mesh of
$100\times50\times50$, with $nt=12$ steps per year, in a $Pentium(R)IV$ processor
(3.2 GHz, 1Gb RAM). Results are quite satisfactory.

\begin{center}
\begin{tabular}{|c|c|c|c|}
\hline
TMc & Tc & Ntotal & CpuTime \\
\hline
2   &  1 &  12 & 0.9 s\\
\hline
11   &  10 & 120 & 7.3 s\\
\hline
20   &  19 & 228 & 14.0 s\\
\hline
\end{tabular}
\end{center}

For illustration,   the numerical solutions obtained for the Caplet's prime and the greeks $\rho=\frac{\partial C}{\partial r}$ and
$vega, \nu=\frac{\partial C}{\partial v}$ are shown in figures (\ref{fig17},\ref{fig18},\ref{fig19}) respectively. The computations
have been performed  for  TMc=2, Tc=1 and Ntotal=12 and the pictures are shown at y=0.

\section{Some additional numerical aspects}
\label{additional}

\subsection{Metric choice} As commented in section \ref{metrics}, it is highly
recommendable to introduce a metric layer in the numerical method,
such that the domain under study transforms into a computational
domain which is typically easier to handle (in most cases, this
one is the unity cube), as long as this domain enables the use of
structures uniform meshes, where one can concentrate the mesh
points wherever needed. The mesh that has been used in this work
is hyperbolic (see eq. \ref{metrica}), following Tavella
\cite{Tavella}. One of the main properties of these meshes is that
one can concentrate as many points as needed in the inner regions
of the zone under study, in order to achieve a better resolution.
It is thus convenient to fix the parameters related to the domain
transformation. For instance, $r$ will transform according to:

$$
r = K_r + \alpha_r \sinh(c_{2r} x_r + c_{1r}(1 - x_r)),
$$
$$
 c_{1r} = \mathrm{asinh}\left(\frac{(r_{0}-K_r)}{\alpha_r}\right),
\quad
c_{2r}=\mathrm{asinh}\left(\frac{(r_{\infty}-K_r)}{\alpha_r}\right),
$$

where the jacobian of the transformation reads:

$$
\frac{dr}{dx_r}=\alpha_r \sinh(c_2 x_r + c_1(1 -
x_r))(c_{2r}-c_{1r}).
$$

\begin{figure}[htpb]
\begin{center}
 \includegraphics[width=10cm,height=10cm]{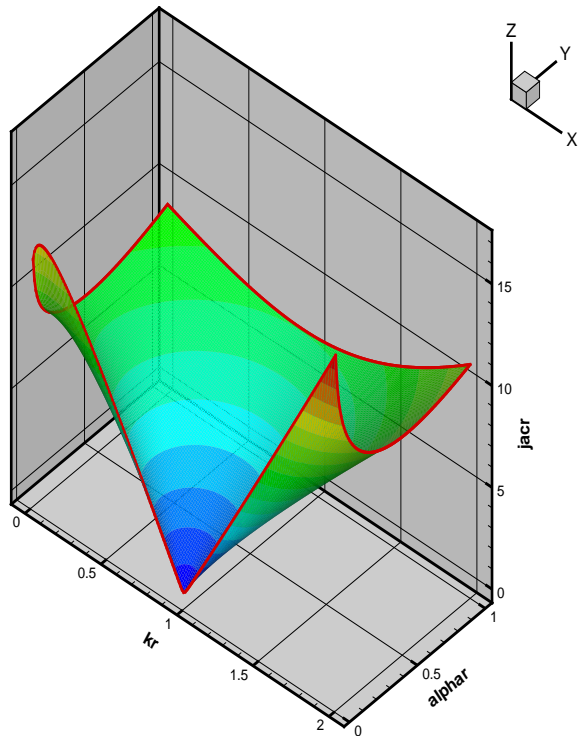}
\caption{Variation of the transformation's Jacobian in $r=1$ as a
function of $K_r$ and $\alpha_r$.}
\label{fig14}
\end{center}
\end{figure}

We have then four parameters $K_r, \alpha_r, r_{0}, r_{\infty}$ to
fix:
\begin{itemize}
\item {\bf  $ r_{0}, r_{\infty}$} define the real domain of study.
Obviously $r_{0}=0$. On the other side, $r_{\infty}$ must be such
that his values doesn't modify the solution in the zone under
study (that is, close to the strike). There is no recipe in order
to find the adequate value, but after some preliminary estimations
and taking into account the boundary conditions, we have set
$r_\infty = 250$.

\item {\bf $K_r$} defines the region under study, that is, a
neighborhood of the strike.

\item {\bf $\alpha_r$} This parameter provides a measure of the
mesh's stretching, i.e. the number of points that will be
concentrated in the zone of interest -close to the strike-.
Concretely, the smaller $\alpha_r$, the larger concentration.
Given that its value also affects the jacobian of the
transformation, it is desirable that $\alpha_r$ is such that the
jacobian be close to 1. In figure (\ref{fig14}) we plot this dependence, for
$r=1 (r_0)$. Note that for $\alpha_r \simeq 0.05-0.1$, the
jacobian reaches the unity. A similar study for $v$ e $y$ lead us
to fix $\alpha_v \simeq 0.5$ y $\alpha_y\simeq 0.05$.
\end{itemize}

As a summary, the metric's characteristic values are:

$$
r_0=v_0=y_0=0, \ \ r_\infty=y_\infty=250, v_\infty=30,
$$

$$
\alpha_r=\alpha_y=0.05, \ \ K_r = Strike, K_y=0, K_v= input
(\simeq 0.5)
$$

\subsection{Softening of the initial condition}

\begin{figure}[htpb]
\begin{center}
 \includegraphics[width=8cm,height=8cm]{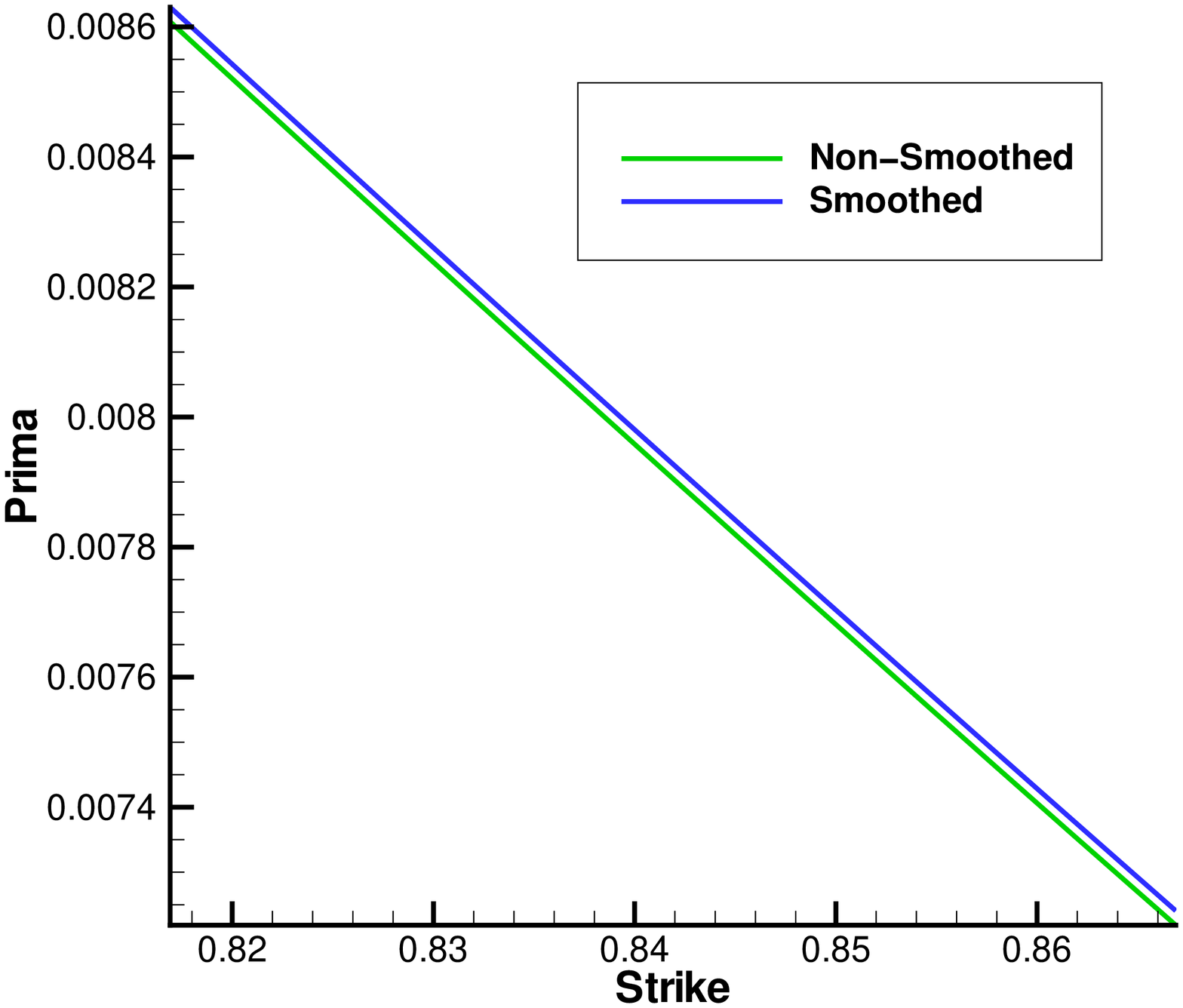}
\caption{Premium variation for a generic Caplet and different
strikes. Simulations have been realized in a single mesh of
100x40x40 with nt=12 steps per year.}
\label{fig15}
\end{center}
\end{figure}

\begin{figure}[htpb]
\begin{center}
  \includegraphics[width=8cm,height=8cm]{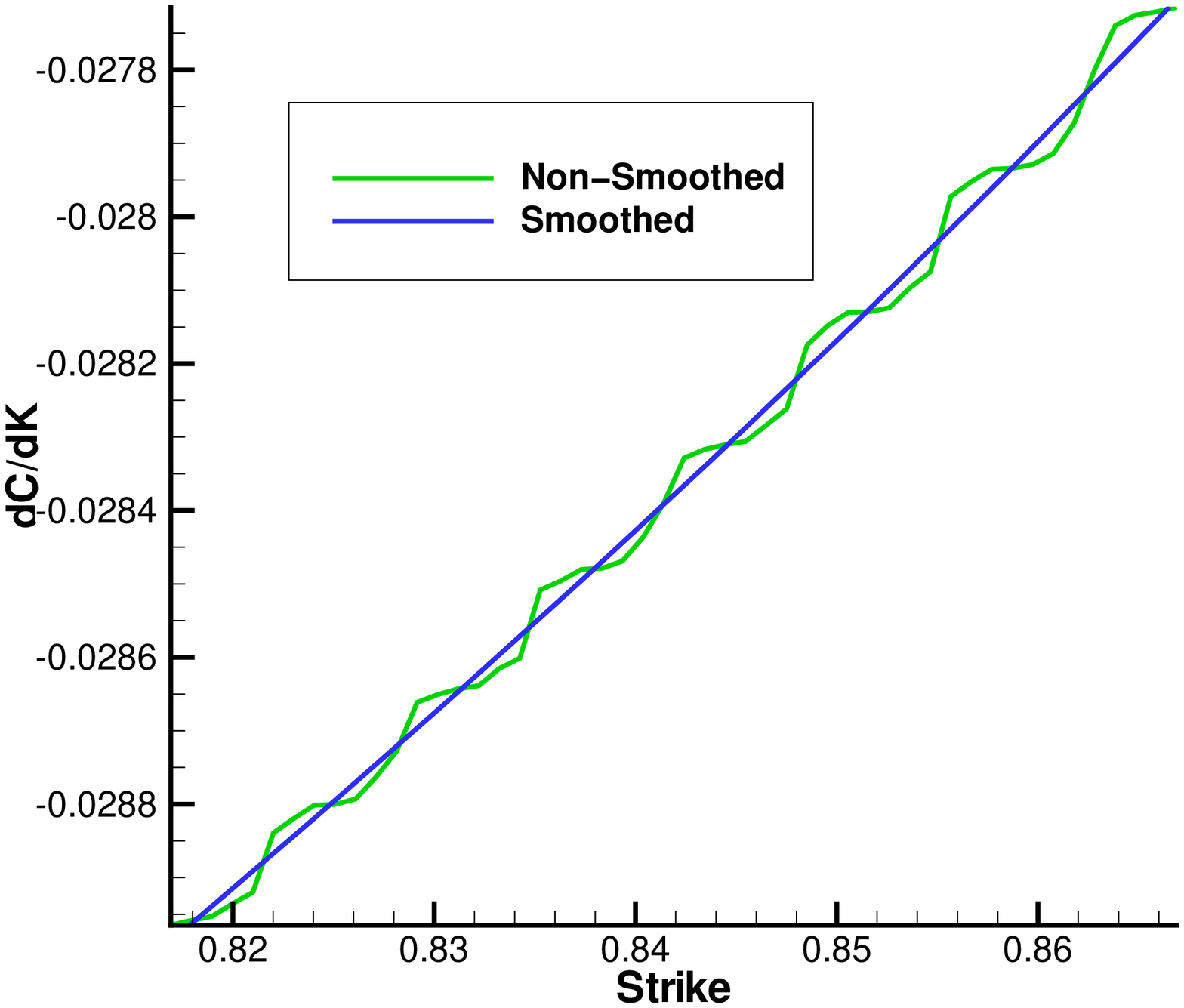}
  \caption{Variation of the premium derivative for a generic
Caplet and different strikes. Simulations have been realized in a
single mesh of 100x40x40 with nt=12 steps per year.}
\label{fig16}
\end{center}
\end{figure}

\begin{figure}[htpb]
\begin{center}
  \includegraphics[width=8cm,height=8cm]{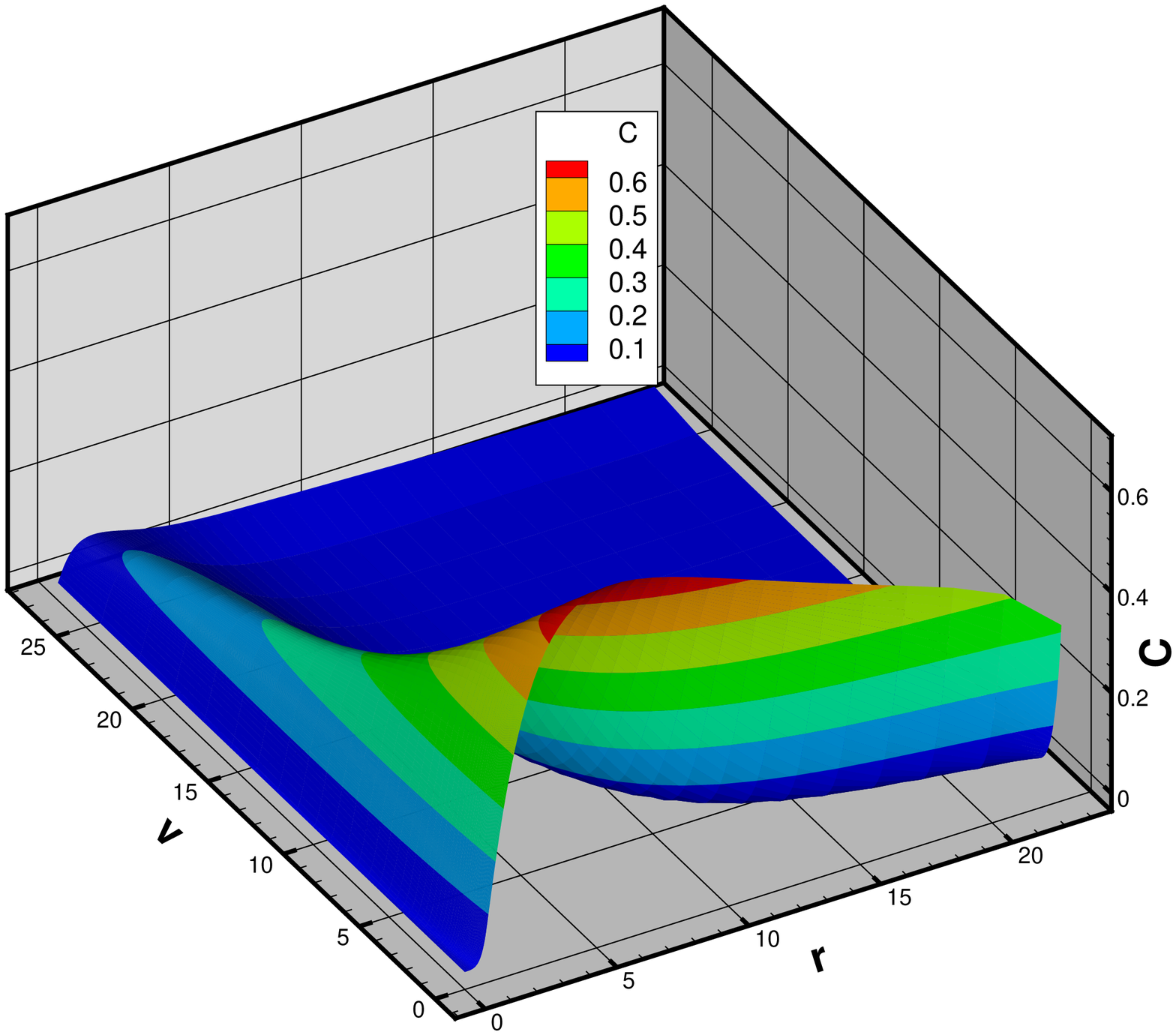}
  \caption{3-Dimensional view of the premium for a generic Caplet at y=0. Simulations have been realized in a
single mesh of 100x50x50 with nt=12 steps per year.}
\label{fig17}
\end{center}
\end{figure}

\begin{figure}[htpb]
\begin{center}
  \includegraphics[width=8cm,height=8cm]{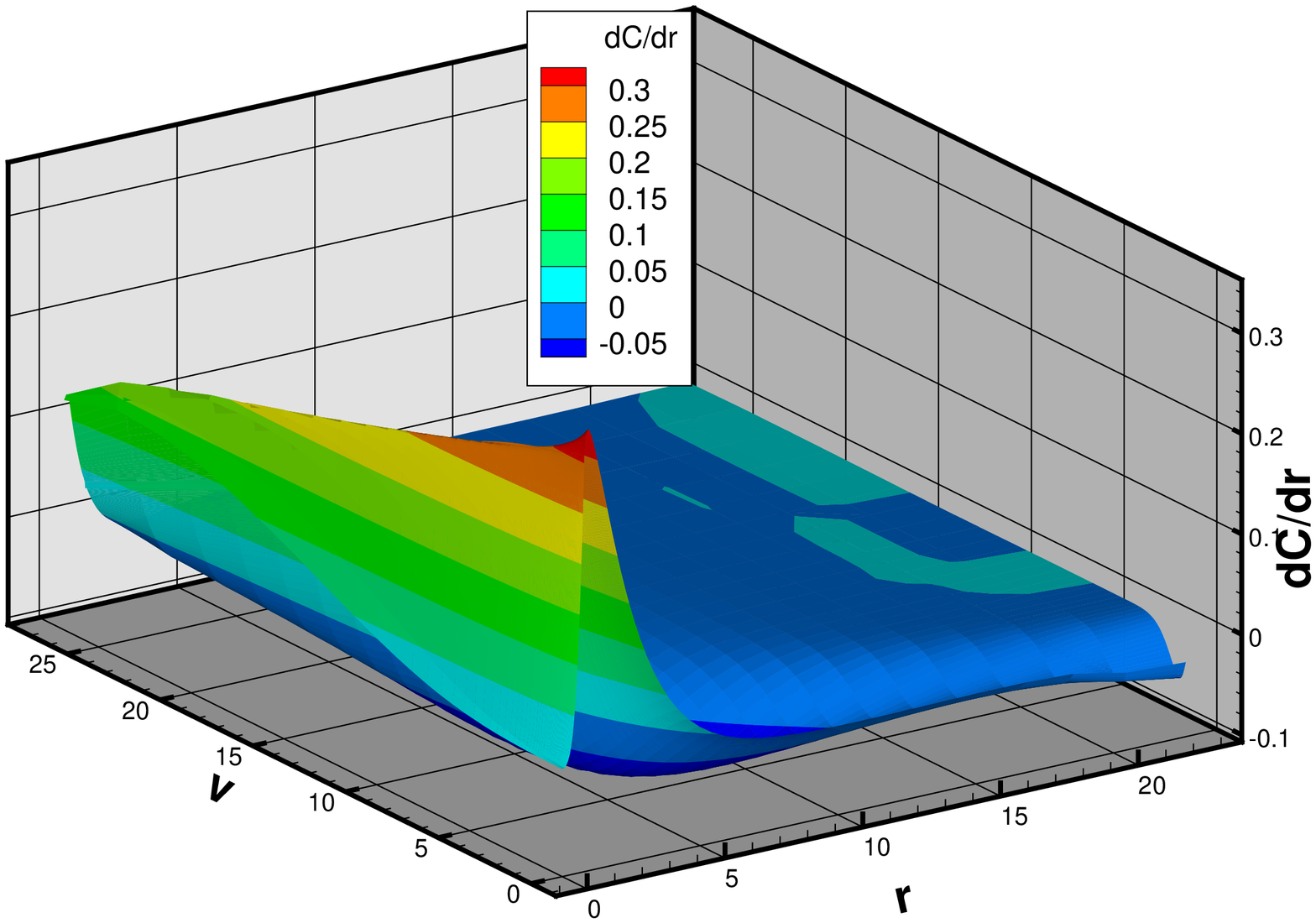}
  \caption{3-Dimensional view of  $\rho$ for a generic Caplet at y=0. Simulations have been realized in a
single mesh of 100x50x50 with nt=12 steps per year.}
\label{fig18}
\end{center}
\end{figure}

\begin{figure}[htpb]
\begin{center}
  \includegraphics[width=8cm,height=8cm]{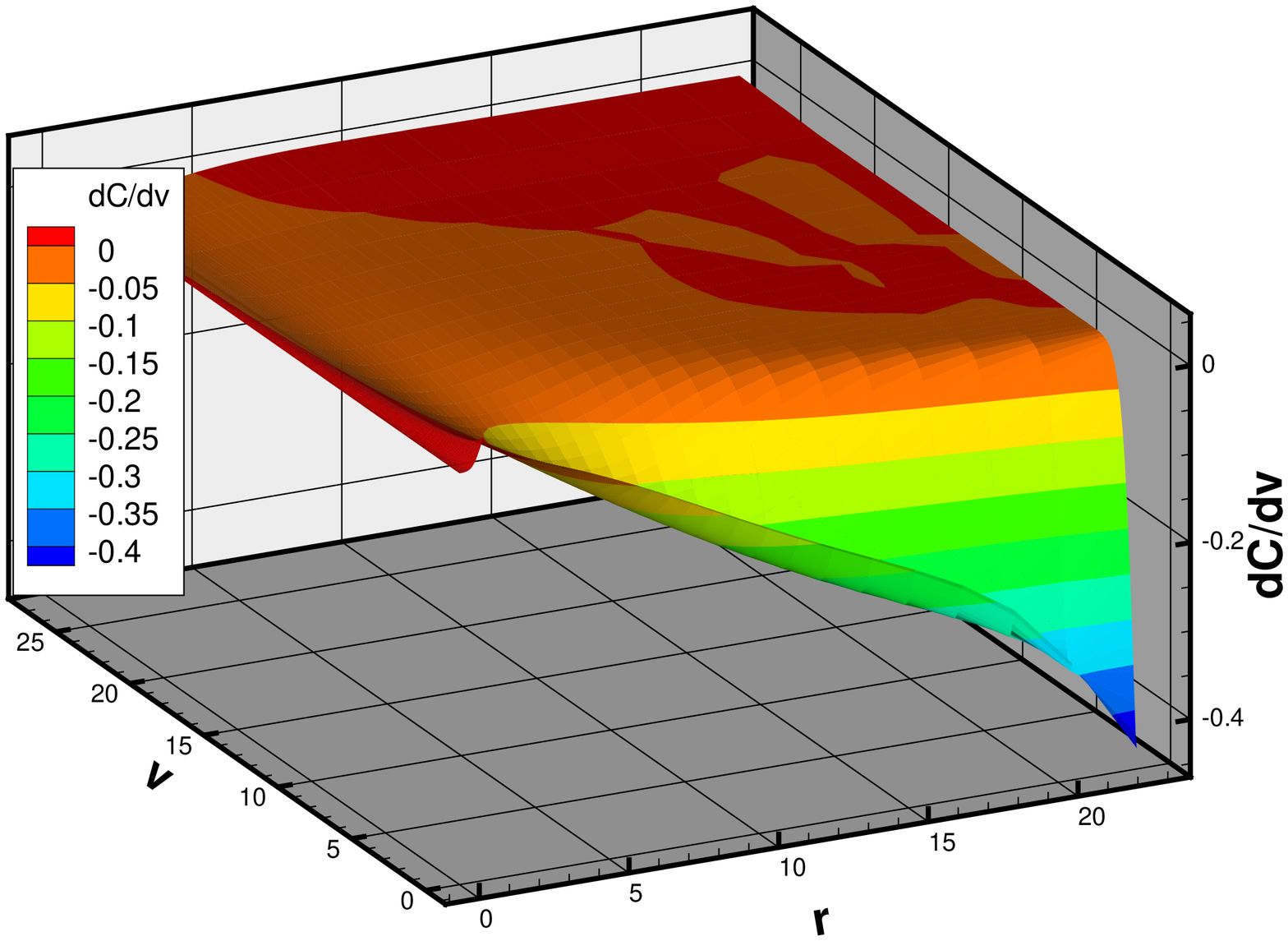}
  \caption{3-Dimensional view of  $vega$ for a generic Caplet at y=0. Simulations have been realized in a
single mesh of 100x50x50 with nt=12 steps per year.}
\label{fig19}
\end{center}
\end{figure}

Following Tavella \cite{Tavella}, as far as the payoff is
typically a discontinuous function, small variations in the strike
lead to a non smooth behavior of the solution. This is not
desirable and therefore some numerical techniques should be
applied in order to soften it:
\begin{itemize}
\item Perform a dynamical modification of the mesh, related to the
payoff's shape. This is an elegant solution, however for practical
purposes this technique is not well fitted as long as it usually
leaves to mesh interpolation.
\item Soften the initial/final
conditions. In order to do so, one can define an average
initial/final condition in the following terms:

$$ C_{ijk}(t,r,v,y)=\frac{1}{\omega}\int_\omega C(t,r,v,y) d\omega, $$
where $\omega$ is a control element centered in the mesh point
$r_i,v_j,y_k$. The effect of this average is represented in
figures (\ref{fig15}-\ref{fig16}).

\end{itemize}

\subsection{Boundary condition for variable $y$} Note that the
HJM model under hands is only convective for variable $y$ (first
derivatives are null for every variable but $y$). When $y=0$, the
solution's characteristic crosses the domain, what indicates that
the boundary takes some information from inside. Consequently, the
discretization of both the interior and the boundary should be
consistent, and then a second order scheme should be applied to
the boundary discretization. We have implemented two different
possibilities in the numerical scheme:

\begin{itemize}
\item Advanced first order differences:
$$
\frac{\partial C}{\partial y} \simeq
\frac{C^n_{i,j,1}-C^n_{i,j,0}}{\Delta y}
$$
\item Advanced second order differences:
$$
\frac{\partial C}{\partial y} \simeq \frac{ -C^n_{i,j,2} + 4
C^n_{i,j,1} - 3 C^n_{i,j,0}}{2 \Delta y}.
$$
\end{itemize}
and quite surprisingly, no significative differences have been
found between both schemes results.

\section{Conclusions}
\label{conclusion}
 In this paper we have proposed a complete numerical methodology to efficiently solve a
single factor HJM model with stochastic
volatility. For this task we have first Markovianised and reformulated the
model in terms of a three dimensional PDE system. The numerical method involved finite-difference and Crank-Nicholson schemes for the spatial
and temporal discretization respectively. In order to
decrease the computing time without loosing precision, we have
successfully applied ADI schemes and performed a preliminary scale analysis to optimize the mesh resolution.
The validation of the numerical schemes has been done comparing the numerical solution of some
test curves (zero coupon bond, Caplet) with analytical models and
a Heston model respectively. The goodness of the results in terms of low computation time (order of seconds in a standard pc)
and good accuracy (typical errors obtained either
for artificial or quite realistic forward interest rate curves
haven't gone beyond $10^{-4}$ in any case) suggest that the method is suitable to be applied in
realistic applications, concretely in the financial industry.\\

\noindent \textbf{Acknowledgments} The authors thank anonymous referees for their helpful suggestions. LL acknowledges
financial support from grants FIS2009-13690 and S2009ESP-1691.

\bibliographystyle{model1-num-names}

\section{Figures}

\newpage

\end{document}